\documentclass[11pt,english,longbibliography]{article}
\usepackage{jheppub}

\usepackage{amsthm}

\makeatletter
\newcommand*{\rom}[1]{\expandafter\@slowromancap\romannumeral #1@}
\makeatother

\usepackage{verbatim}
\usepackage{prettyref}
\usepackage[numbers,sort&compress]{natbib}
\usepackage{amsmath}
\usepackage{amssymb}
\usepackage{tikz-cd}
\tikzset{commutative diagrams/row sep/huge=4cm}
\tikzset{commutative diagrams/column sep/huge=4cm}
\tikzcdset{scale cd/.style={every label/.append style={scale=#1},
    cells={nodes={scale=#1}}}}
\usepackage{lmodern}
\usepackage{tcolorbox}
\usepackage{cases}
\usepackage{bbold}
\usepackage{bm}
\usepackage{ytableau}
\usepackage{youngtab}
\usepackage{mathtools}
\usepackage{graphicx}
\usepackage[nottoc]{tocbibind}
\usepackage{mathrsfs}
\usepackage{caption}
\usepackage{subcaption}
\usepackage{cancel}
\usepackage{float}
\usepackage{tikz}
\usepackage{slashed}
\usepackage{braket}

\newcommand{\tr}{\operatorname{tr}}

\usetikzlibrary{arrows.meta}
\usetikzlibrary{bending}
\usepackage{tikz}
\usepackage{tikz}
\usetikzlibrary{decorations.pathmorphing,calc}
\tikzset{
    Witten diagram/.style={
        execute at begin picture={
            \draw[blue, line width=1.5pt] circle[radius=\pgfkeysvalueof{/tikz/Witten/radius}];
            \path node (X){\phantom{X}};
        },
        baseline={(X.base)}
    },
    vertex/.style={circle,fill,inner sep=1.5pt,node contents={}},
    Witten/.cd,
    radius/.initial=3cm
}
\usetikzlibrary{patterns}
\usepackage{lipsum}
\usepackage{framed}
\usepackage{adjustbox}
\usepackage{mathtools}

 \usepackage{slashed}
\usepackage{esvect}
\usepackage{dsfont}

\usepackage{color}
\definecolor{darkgreen}{rgb}{0,0.5,0}
\definecolor{darkblue}{rgb}{0,0,0.6}
\definecolor{purple}{rgb}{0.4,.2,0.7}

\numberwithin{equation}{section}
\numberwithin{figure}{section}
\numberwithin{table}{section}

\usepackage{comment}

\newcommand{\affsqrt}{\ensuremath{\sqrt{\epsilon}}}
\newcommand{\affeps}{\ensuremath{\epsilon}}
\newcommand{\affepsfive}{\ensuremath{\epsilon^{\frac{5}{4}}}}
\title{\boldmath $\mathcal{PT}$-symmetric Field Theories at Finite Temperature}
\author[\affsqrt]{Oleksandr Diatlyk}
\author[\affeps]{Andrei Katsevich}
\author[\affepsfive]{Fedor K. Popov}
\affiliation[\affsqrt]{Center for Cosmology and Particle Physics, New York University, New York, NY 10003, USA}
\affiliation[\affeps]{Joseph Henry Laboratories, Princeton University, Princeton, NJ 08544}
\affiliation[\affepsfive]{Simons Center for Geometry and Physics, Stony Brook University, Stony Brook, NY}

\abstract{We investigate the thermal properties of \(\mathcal{PT}\)-symmetric scalar field theories with purely imaginary couplings. The free energy governs the asymptotic density of states, providing an effective measure of the number of degrees of freedom, while thermal masses and one-point functions provide predictions for operator dimensions and three-point functions in the corresponding \(d=2\) Conformal Field Theories. Naive finite-temperature perturbation theory near upper critical dimensions is spoiled by infrared divergences. To remove these divergences, we introduce a ``thermal normal-ordering'' scheme that resums these contributions and yields a systematic \(\epsilon\)-expansion. This framework allows us to compute the free energy, thermal masses, and one-point functions in the cubic and quintic \(O(N)\) models. We compare the thermal free energy density, thermal masses, and one-point function in two dimensions with exact results derived from the proposed Ginzburg--Landau descriptions of the non-unitary minimal models \(M(2,5)\) and \(M(3,8)_D\). Eventually, we employ two-sided Pad\'e extrapolations to obtain estimates for the thermal free energy in $d=3,4,5$.}

\begin{document} 
\maketitle
\newpage
\section{Introduction}
\label{sec:Introduction}

Recently, there has been a surge of interest in non-unitary Conformal Field Theories (CFTs). 
Due to unbroken $\mathcal{PT}$ symmetry \cite{Bender:2018pbv}, these theories possess a real spectrum.
A central question in the study of any CFT concerns the spectrum of operators $\mathcal{O}_{n,s}$ and their scaling dimensions $\Delta_{n,s}$. This spectral data is crucial as it governs various experimental observables; however, its computation remains notoriously difficult due to the strongly coupled origin of the problem, even for the lightest operators in the spectrum.

A powerful approach to gain insights into the asymptotic behavior of the spectrum is to formulate the theory on distinct manifolds. The partition function on such manifolds can then be related to a sum over quantities that are directly computable from the Hilbert space of the underlying CFT. For instance, if we compute the partition function of a CFT on the $S^1_\beta\times S^{d-1}$, where the sphere $S^{d-1}$ has radius $R$ and the thermal circle $S^1_\beta$ has a circumference equal to the inverse temperature $\beta$, we can immediately infer the following relation:
\begin{equation}
    Z_{S_\beta^1\times S^{d-1}}(\beta) = \sum_{n,s} e^{-\frac{\beta}{R} \Delta_{n,s}} = \int d\Delta \rho(\Delta) e^{-\frac{\beta}{R}\Delta}\,.
\end{equation}
Thus, by performing an inverse Laplace transform, one could recover the spectrum. While an exact computation is often intractable, taking the high-temperature limit $\frac{\beta}{R} \to 0$ allows us to argue that the partition function behaves as:
\begin{equation}
    \log Z_{S_\beta^1\times S^{d-1}}= - \beta A_{d-1}R^{d-1}f + \mathcal{O} \left(\frac{R^{d-3}}{\beta^{d-3}}\right)\,.
\end{equation}
Here, $A_{d-1}=\frac{2\pi^{\frac{d}{2}}}{\Gamma(\frac{d}{2})}$ is the surface area of the unit $(d-1)$-dimensional sphere, $f$ represents the free energy density at inverse temperature $\beta$ and is fixed as a function of $\beta$ by a conformal symmetry up to a coefficient. We used a simple physical argument: for a large radius $R \gg \beta$, the system effectively resides in a flat space-time of volume $\beta A_{d-1} R^{d-1}$. Performing an inverse Laplace transform, we find that the asymptotic density of states behaves as:
\begin{equation}
\begin{aligned}
    \log \rho(\Delta) \propto d \left(\frac{\Delta}{d-1}\right)^\frac{d-1}{d}(-\beta^df A_{d-1})^\frac{1}{d}\,.
\end{aligned}
\end{equation}
This result shows that studying the thermal free energy allows us to directly determine the asymptotic behavior of the density of states.

Analogously, we can compute the thermal one-point function of a local operator $\phi$ in the geometry $S^1_\beta\times S^{d-1}$. By expanding the trace over the Hilbert space, we express the expectation value as a sum over the operator spectrum, weighted by the diagonal Operator Product Expansion (OPE) coefficients $C_{n\phi n}$ (or equivalently, the diagonal matrix elements $\braket{n|\phi|n}$). In the high-temperature limit $\frac{\beta}{R} \to 0$, conformal invariance dictates that the expectation value must scale with temperature as $T^{\Delta_\phi}$. Matching the spectral sum to this universal scaling behavior yields:
\begin{equation}
\begin{aligned}
    \braket{\phi}_{S_\beta^1\times S^{d-1}} = \frac{1}{R^{\Delta_\phi}} \frac{\sum\limits_n C_{n\phi n} e^{-\frac{\beta}{R}\Delta_n}}{\sum\limits_n e^{-\frac{\beta}{R} \Delta_n}} \xrightarrow{\frac{\beta}{R} \to 0} \frac{a_\phi}{\beta^{\Delta_\phi}}.
\end{aligned}
\end{equation}
From this relation, one can deduce the asymptotic behavior of the average diagonal matrix elements for heavy states ($\Delta \gg 1$). Assuming the validity of the equivalence between the microcanonical and canonical ensembles, the coefficients must satisfy:
\begin{equation}
    C_{n\phi n} \approx a_\phi \left( \frac{ \Delta_n}{(d-1)(-\beta^df)A_{d-1}}  \right)^{\frac{\Delta_\phi}{d}},
\end{equation}
where $a_\phi$ is a theory-dependent normalization constant, and thus we can also extract the asymptotic behavior of  OPE coefficients.

Crucially, for $\mathcal{PT}$-symmetric field theories, this thermodynamic procedure remains a valid method for counting the asymptotic density of states. To demonstrate this, consider a Hamiltonian $H$ diagonalized in a biorthogonal basis:
\begin{equation}
\begin{aligned}
    &H = \sum E_n \ket{n_R}\bra{n_L}, \quad \braket{n_L|m_R} = \delta_{nm}, \quad  \sum_n \ket{n_R}\bra{n_L} = \mathbb{1},\quad \ket{n_R} \neq \bra{n_L}^\dagger\,, \\
    &\tr(e^{-\beta H})
    = \sum^\infty_{k=0} \frac{(-\beta)^k}{k!}  \sum_n E_n^k = \sum_n e^{-\beta E_n} = e^{-\beta F}\,.
\end{aligned}
\end{equation}
By the operator-state correspondence, we therefore expect the partition function of a $\mathcal{PT}$-symmetric field theory on $S_\beta^1\times S^{d-1}$ to compute this trace precisely. Thus, we expect the thermal free energy to be a real, negative number that controls the asymptotic spectrum. Analogously, by computing the one-point function of operators, we can extract the asymptotic behavior of OPE coefficients in these theories. 

Another interesting observable is the thermal mass \(m_{\rm th}\), which determines the exponential decay of the finite-temperature two-point function at large spatial separation,
\begin{equation}
    \langle \phi(x)\phi(0)\rangle_\beta \sim e^{-m_{\rm th}|x|}\,.
\end{equation}
In two dimensions, the same observable can be computed by quantizing the theory on \(S^1_\beta\times \mathbb{R}\). For instance, the exponential decay along \(\mathbb{R}\) is governed by the energy gaps of CFT on the circle, i.e., dimensions of local operators:
\begin{equation}
    \langle \phi(x)\phi(0)\rangle_\beta = \sum_{n}\braket{\Omega_L|\phi|n_R}\braket{n_L|\phi|\Omega_R}\exp\left(-\frac{2\pi\Delta_{\text{eff},n}}{\beta}|x|\right)\,,
\end{equation}
where \(\ket{\Omega_R}\) denotes the ground state of CFT on the circle and \(\Delta_{\text{eff},n}\equiv\Delta_n-\Delta_0>0\) are the gaps between the $n$-th excited state and true ground state of CFT on \(S^1_\beta\), that for non-unitary theories could be different from the identity operator. Therefore the dimensionless combination \(  m_{\rm th} \beta\) provides a direct prediction for the effective scaling dimension in \(d=2\)\footnote{Analogous relation was obtained in \cite{Itzykson:1986pk} for the Yang-Lee model}:
\begin{equation}\label{eq:Deltaeff}
    m_{\rm th}  \beta=2\pi\Delta_{\text{eff}}\,,
\end{equation}
where $\Delta_{\text{eff}}=\Delta_{\text{eff},n}$ denotes the gap between the ground state and the lowest excited state $\ket{n_R}$ whose OPE coefficient with $\phi$ is nonzero, namely $\braket{\Omega_L|\phi|n_R} \propto C_{\Omega\phi n}\neq0$. By analytically continuing the \(6-\epsilon\) result and formally setting \(\epsilon=4\), we thus obtain a prediction for \(\Delta_{\text{eff},n}\) of the corresponding two-dimensional non-unitary CFT. For unitary theories with $\Omega=I$ \eqref{eq:Deltaeff} gives known $ m_{\rm th} \beta =2\pi\Delta_{\phi}$ \cite{cardy1984conformal}.

The thermal one-point function admits a similarly direct interpretation. In contrast to unitary theories, where one-point functions of nontrivial primaries on the cylinder vanish because the ground state on \(S^1_\beta\) is created by the identity operator, in the non-unitary theories considered here, the ground state on \(S^1_\beta\) is created by a nontrivial primary $\Omega$. As a result, the thermal one-point function is generically nonzero in these theories. Quantizing on \(S^1_\beta\times \mathbb{R}\), one finds
\begin{equation}\label{eq:1ptfunction2D}
    \langle \phi\rangle_\beta
    =
    \left(\frac{2\pi}{\beta}\right)^{\Delta_\phi} C_{\Omega\phi\Omega}\,.
\end{equation}
Comparing this with the high-temperature form \(\langle \phi\rangle_\beta=a_\phi \beta^{-\Delta_\phi}\), we obtain
\begin{equation}\label{eq:OPE1pt}
    a_\phi=(2\pi)^{\Delta_\phi}C_{\Omega\phi\Omega}\,.
\end{equation}
Thus, by analytically continuing the \(6-\epsilon\) result for \(a_\phi\) to \(\epsilon=4\), we obtain a prediction for the corresponding two-dimensional diagonal OPE coefficient. In the cases of interest, where \(\Omega=\phi\), this reduces to a prediction for \(C_{\phi\phi\phi}\).

A fundamental challenge in modern theoretical physics is identifying a quantity that measures the number of degrees of freedom and monotonically decreases along the renormalization group (RG) flow. In $d=2$ and $d=4$, this is addressed by the powerful $c$- and $a$-theorems, which strictly constrain the flow \cite{Zamolodchikov:1986gt, Cardy:1988cwa,Osborn:1989td,Jack:1990eb,Komargodski:2011vj}. For unitary theories in an arbitrary dimension $d$, a generalized $F$-theorem was proposed \cite{Jafferis:2011zi,Klebanov:2011gs,Giombi:2014xxa,Fei:2015oha}, governed by the generalized sphere free energy $\tilde{F} \equiv \sin(\frac{\pi d}{2}) \log Z_{S^d}$, which naturally reduces to the central charge $c$ in two dimensions. However, for non-unitary flows, the standard $c$-, $a$-, and $F$-theorems are generally violated \cite{Fei:2015oha,Giombi:2024zrt}. Nonetheless, in $d=2$, a $c_{\text{eff}}$-theorem successfully generalizes the standard $c$-theorem to a class of flows between non-unitary $\mathcal{PT}$-symmetric field theories \cite{Itzykson:1986pk,Castro-Alvaredo:2017udm}. An obvious open question is whether one can formulate an analogous $F_{\text{eff}}$-theorem for higher dimensions to constrain non-unitary flows similarly.

Since free energy at finite temperatures is also a measure of degrees of freedom, it is natural to formulate an analogous $c_{\text{Therm}}$-theorem, where $c_{\text{Therm}}$ is normalized thermal free energy:
\begin{equation}\label{eq:freeenergydensity}
   f=-\frac{\Gamma(\frac{d}{2})\zeta(d)}{\pi^{\frac{d}{2}}}c_{\text{Therm}}T^d\,.
\end{equation}
For a real free scalar boson, $c_{\text{Therm}}=1$. However, there are well-known counterexamples to this proposal; for example, the three-dimensional flow from the quartic $O(N)$ model to $N-1$ free Goldstone bosons \cite{Sachdev:1993pr,Chubukov:1993aau}. Nonetheless, in $d=2$, $c_{\text{Therm}}$ is the effective central charge $c_\text{eff}$. Thus, $c_{\text{Therm}}$ is a natural generalization of the effective central charge $c_\text{eff}$ to any dimension. In this paper, we test the $c_{\text{Therm}}$-theorem for the flow between two non-unitary fixed points of the $N=1$ cubic model, in which the $F$-theorem is violated \cite{Giombi:2024zrt}. Another major advantage of thermal free energy, compared with sphere free energy, is that it can be analytically continued down to $d=2$, where it can be directly compared with known exact solution.

Another motivation for studying the thermal free energy of conformal field theories is the recently appeared conjectures for Ginzburg-Landau (GL) descriptions of some classes of non-unitary minimal models $M(p,q)$ \cite{Klebanov:2022syt,Katsevich:2024jgq,Katsevich:2025ojk}. In \(d=2\), they have effective central charge \cite{Itzykson:1986pk}:
\begin{equation}\label{eq:ceff}
    c_\text{eff}(p,q)=1-\frac{6}{pq}\,.
\end{equation}
Analytically continuing the thermal free energy to \(d=2\) and using (\ref{eq:freeenergydensity}) yields \(c_\text{eff}(p,q)\), enabling a direct test of the GL conjectures \cite{Zamolodchikov:1986db,Cardy:1985yy,Klebanov:2022syt,Katsevich:2024jgq,Katsevich:2025ojk}.

Let us consider cubic $O(N)$ model introduced in \cite{Fei:2014yja} with the action:
\begin{equation}
    S_\text{cubic}=\int d^dx\left(\frac{1}{2}(\partial_\mu\phi_i)^2+\frac{1}{2}(\partial_\mu\sigma)^2+\frac{g_1\sigma\phi_i^2}{2}+\frac{g_2\sigma^3}{3!}\right)\,.
\end{equation}
This theory has both a $\mathbb{Z}_2$ symmetry, acting as $\phi_i\rightarrow-\phi_i$, and a $\mathcal{PT}$ symmetry, acting as $\sigma\rightarrow-\sigma$ and $i\rightarrow-i$. Previous studies have demonstrated the existence of unitary stable fixed points for $N>1038$ \cite{Fei:2014yja,Fei:2014xta,Gracey:2015tta,Kompaniets:2021hwg}. Furthermore, in the large $N$ limit this theory coincides with the large $N$ critical $O(N)$ model in $d=6-\epsilon$. In addition, there exist non-unitary stable fixed points for $N<1.02$ with purely imaginary couplings. The renormalization of the cubic $O(N)$ model in dimensional regularization has been carried out up to five loops \cite{Fei:2014xta,Gracey:2015tta,Kompaniets:2021hwg}. This model has also been studied using the Functional Renormalization Group (FRG) \cite{Mati:2014xma,Mati:2016wjn,Eichhorn:2016hdi,Kamikado:2016dvw,Connelly:2020gwa}.

The $N=0$ fixed point describes the Yang--Lee universality class \cite{Fisher:1978pf}:
\begin{equation}\label{eq:YLaction}
    S_\text{YL}=\int d^dx\left(\frac{1}{2}(\partial_\mu\phi)^2+\frac{g\phi^3}{3!}\right),\quad g\in i\mathbb{R}\,.
\end{equation}
It has only $\mathcal{PT}$ symmetry, under which $\phi\rightarrow-\phi$ and $i\rightarrow-i$. It first appeared as an effective description of the accumulation of zeros of the partition function of the Ising model in an imaginary magnetic field. In $d=2$, the $N=0$ stable fixed point describes the simplest non-unitary minimal model $M(2,5)$ \cite{Cardy:1985yy}. A Lagrangian formulation of the Yang--Lee universality class was developed in \cite{Fisher:1978pf}, enabling systematic computations, and its \(\epsilon\)-expansion has been pushed to six loops \cite{Gracey:2015tta,Kompaniets:2021hwg,Borinsky:2021jdb,Schnetz:2025wtu,Gracey:2025rnz}. The Yang--Lee model has also been studied using nonperturbative methods such as the FRG \cite{An:2016lni,Zambelli:2016cbw,Rennecke:2022ohx,Benedetti:2026tpa}, high-temperature expansions \cite{Butera:2012tq}, the non-unitary bootstrap \cite{Gliozzi:2013ysa,Gliozzi:2014jsa,Hikami:2017hwv}, and fuzzy-sphere regularization \cite{ArguelloCruz:2025zuq,Fan:2025bhc,EliasMiro:2025msj}.

Also, there is the $N=1$ stable fixed point \cite{Fei:2014xta}:
\begin{equation}
    S_{3,8}^D=\int d^dx\left(\frac{1}{2}(\partial_\mu\phi)^2+\frac{1}{2}(\partial_\mu\sigma)^2+\frac{g_1\sigma\phi^2}{2}+\frac{g_2\sigma^3}{3!}\right),\quad g_1,g_2\in i\mathbb{R}\,.
\end{equation}
In $d=2$, it gives a GL description of $M(3,8)_D$ minimal model \cite{Fei:2014xta,Klebanov:2022syt,Katsevich:2024jgq}. Additionally, there is an unstable $N=1$ fixed point  $g_{1,\star}=g_{2,\star}$, that could be recasted in the form of the product of two YL theories $M(3,10)_D=M(2,5)\otimes M(2,5)$ \cite{Kausch:1996vq,Quella:2006de,Ardonne_2011}. 

Another $\mathcal{PT}$-symmetric theory which can be considered is the quintic $O(N)$ model. The $N=0$ fixed point describes the Tricritical Yang-Lee (TYL) universality class \cite{Lencses:2022ira,Katsevich:2025ojk}:
\begin{equation}
    S_{\text{TYL}}=\int d^dx\left(\frac{1}{2}(\partial_\mu\phi)^2+\frac{g\phi^5}{5!}\right),\quad g\in i\mathbb{R}\,.
\end{equation}
Recently, it was conjectured that the quintic model describes the non-unitary $M(2,7)$ minimal model in $d=2$ \cite{Katsevich:2025ojk}. The $\epsilon$-expansion of the $i\phi^5$ field theory is currently available only to leading order \cite{Codello:2017epp,Gracey:2017okb}. The quintic model has also been studied using the FRG \cite{Zambelli:2016cbw,Codello:2017epp,Benedetti:2026tpa}.

In this work, we investigate the aforementioned $\mathcal{PT}$-symmetric field theories using the $\epsilon$-expansion. The direct computation of the free energy is plagued by infrared divergences, which become increasingly severe as we consider higher orders of perturbation theory. To derive a systematic $\epsilon$-expansion in the presence of these singularities, we must first resum these divergences. We anticipate that this procedure will yield a finite modification of our parameters, most notably the emergence of a non-zero thermal mass which screens the long-range modes.

A standard approach to this problem is to introduce a small renormalized mass $m^2$ to regulate the theory. One then resums all leading divergences and subsequently takes the limit $m^2 \to 0$. Even though this approach is useful and allows us to compute all interesting quantities in the leading $\epsilon \to 0$ limit, it becomes notoriously convoluted at higher orders of perturbation theory. At these higher orders, nested sub-diagrams are consequently plagued with divergences, thereby complicating the systematic computation within the $\epsilon$-expansion.

Thus, we will strictly limit the use of this regulator technique to a brief computation of the leading-order thermal free energy. We will then immediately abandon it in favor of a more robust framework known as ``thermal normal ordered'' perturbation theory, which systematically handles all divergences appearing at higher orders. The key idea behind this technique is the observation that IR divergences arise due to the existence of non-zero self-contractions of interaction operators. By redefining our operators to eliminate these self-contractions, we induce finite shifts in our fields and naturally generate the requisite thermal mass, rendering the expansion finite.

We obtain the first few terms in thermal free energy for the cubic $O(N)$ model in $\epsilon$ expansion. Using them, we provide a check of Cardy's conjecture of the GL description of $M(2,5)$ minimal model \cite{Cardy:1985yy} and more recent conjecture about the $M(3,8)_D$ model \cite{Fei:2014xta,Klebanov:2022syt,Katsevich:2024jgq}. We perform two-sided Pad\'e resummations for cubic $N=0,1$ models using the exact value for $c_{\text{eff}}$ in $d=2$ as a constraint.

The paper is organized as follows. In Section \ref{sec:LargeN}, we analyze the large $N$ vector model at finite temperature and derive its thermal free energy, which provides a useful point of comparison for the subsequent discussion. In Section \ref{sec:resumationIRdiv}, we examine the breakdown of naive finite-temperature perturbation theory due to infrared divergences and present its explicit resummation, first for the cubic Yang--Lee theory and then for the quintic model. In Section \ref{sec:thermord}, we develop the framework of thermal ordered operators and derive the associated self-consistent gap equations, again illustrating the construction in both cubic and quintic theories. Section \ref{sec:ThermalO(N)} applies this formalism to the cubic $O(N)$ model on $S^1_\beta\times\mathbb{R}^{5-\epsilon}$, where we study the thermal one-point function, the gap equation, and the renormalized free energy. In Section \ref{sec:results}, we specialize these results to several cases of particular interest, including the large $N$ limit, the Yang--Lee theory at $N=0$, and the cubic model at $N=1$, and compare the resulting extrapolations with the proposed two-dimensional minimal-model descriptions. Technical details, including the free energy of a massive scalar field, the small-mass expansion of the relevant sum-integrals, and the perturbative solution of the gap equation, are collected in the appendices.

\section{Large $N$ vector model}\label{sec:LargeN}
In this section, we compute the thermal free energy of the large $N$ vector model at finite temperature $T= \beta^{-1}$. The action of the large $N$ model on $S^1_\beta\times\mathbb{R}^{d-1}$ has the following form:
\begin{equation}
    S_{\text{large }N} = \int d\tau\, d^{d-1}x\left(\frac12 (\partial_\mu \phi_i)^2 + \frac{1}{2} \sigma \phi_i^2 \right)\,,
\end{equation}
where the field $\sigma$ plays the role of a Lagrange multiplier enforcing the constraint $\phi_i^2 = 0$. This theory has $O(N)$ symmetry and $\mathcal{PT}$ symmetry, $i\rightarrow-i$ and $\sigma\rightarrow-\sigma$. In the
large $N$ limit, we can solve this model semiclassically at zero temperature. For instance, we can find two-point functions of $\phi_i$ and $\sigma$:
\begin{equation}\label{eq:2ptfunction}
    \begin{aligned}
    &\braket{\phi_i(x)\phi_j(y)}=\frac{\Gamma(\frac{d}{2}-1)}{4\pi^{\frac{d}{2}}}\frac{\delta_{ij}}{|x-y|^{d-2}}\,,\\
    &\braket{\sigma(x)\sigma(y)}= \frac{C_{\sigma}}{|x-y|^4}\,, \quad C_\sigma=\frac{2^{d+2}\Gamma(\frac{d-1}{2})\sin(\frac{\pi d}{2})}{N\pi^{\frac{3}{2}}\Gamma(\frac{d}{2}-2)}\,.
    \end{aligned}
\end{equation}
At large $N$ limit, we can solve this model semiclassically at any finite temperature. Thus, assuming that $\sigma$ is constant, the thermal free energy density is
\begin{equation}
    f_{\text{large }N}(\sigma) = Nf_\text{free}(\sigma)+\mathcal{O}(N^0)\,, \label{eq:freeenlargeN}
\end{equation}
where the thermal free energy density of a free scalar is given in Appendix \ref{App:FreeEnergyMassive}. We can find the stationary point of (\ref{eq:freeenlargeN}) with respect to $\sigma$. There is a unique real solution to the saddle-point equation for $2<d<4$ and two complex-conjugate solutions in the range $4<d<6$ \cite{Petkou:2018ynm,Giombi:2019upv}, where $\mathcal{PT}$ symmetry is broken. In $d=3$, the thermal free energy $f_{\text{large }N}$ was computed up to order $\mathcal{O}(N^0)$ \cite{Sachdev:1993pr,Chubukov:1993aau,Diatlyk:2023msc}.

In $d=2n+\epsilon$, we can approximate $\sigma$ by expanding in small $\epsilon$. In $d=6+\epsilon$, we obtain the following expansion:
\begin{equation}\label{eq:largeNmass}
    \frac{\sigma_\star}{\pi^2T^2} = \frac{4\sqrt{\epsilon}}{3\sqrt{5}}-\frac{2\epsilon}{3}+\frac{8\epsilon^{\frac{5}{4}}}{3\sqrt{3}5^{\frac{1}{4}}}-\frac{(1-12\gamma_E+1440\zeta'(-3))\epsilon^{\frac{3}{2}}}{18\sqrt{5}}-\frac{2\;5^{\frac{1}{4}}\epsilon^{\frac{7}{4}}}{\sqrt{3}}+\mathcal{O}
    \left(\epsilon^2,\frac{1}{N}\right)\,.
\end{equation}
This stationary point $\sigma_\star$ determines the thermal mass $m_{\text{th},\phi}^2$ of the scalar fields $\phi_i$. It is also related to the thermal one-point function $\braket{\sigma}$, which we compute below in the $\epsilon$ expansion of the cubic $O(N)$ model. To compare with that result, we define a normalized one-point function $\bar{\sigma}_\star$ using the two-point function coefficient in \eqref{eq:2ptfunction}:
\begin{equation}\label{eq:largeN1pt}
\begin{aligned}
    \bar{\sigma}_\star&=\frac{\sigma_\star}{\sqrt{C_\sigma}}=\sqrt{N}\pi^2T^2\bigg(-\frac{i}{3\sqrt{30}}+\frac{i\sqrt{\epsilon}}{6\sqrt{6}}-\frac{\sqrt{2}i\epsilon^{\frac{3}{4}}}{9\;5^{\frac{1}{4}}}\\
    &+\frac{(17-12\gamma_E+1440\zeta'(-3))i\epsilon}{72\sqrt{30}}+\frac{5^{\frac{1}{4}}i\epsilon^{\frac{5}{4}}}{6\sqrt{2}}+\mathcal{O}
    \left(\epsilon^\frac{3}{2},\frac{1}{N}\right)\bigg)\,.
\end{aligned}
\end{equation}
Substituting $\sigma_\star$ back into the functional $f_{\text{large }N}(\sigma_\star)$, we obtain the thermal free energy of the large $N$ vector model:
\begin{equation}\label{eq:largeNON}
\begin{aligned}
    \frac{f_{\text{large }N}^{d=6+\epsilon}}{N\pi^3T^{6+\epsilon}}&=-\frac{2}{945}+\frac{\sqrt{\epsilon}}{405\sqrt{5}}-\frac{13-4\log\left(\pi e^{\gamma_E-2\frac{\zeta'(6)}{\zeta(6)}}\right)}{3780} \epsilon+\frac{4\epsilon^{\frac{5}{4}}}{675\sqrt{3}5^{\frac{1}{4}}}\\
    &-\frac{1-4 \log \left(4\pi e^{2 \gamma_E }\right)+1440\zeta'(-3)}{3240\sqrt{5}} \epsilon^{\frac{3}{2}}
    -\frac{\epsilon^{\frac{7}{4}}}{27 \sqrt{3}5^{\frac{3}{4}}} +\mathcal{O}\left(\epsilon^2,\frac{1}{N}\right)\,.
\end{aligned}
\end{equation}
\section{Resummation of the infrared divergences}\label{sec:resumationIRdiv}
\subsection{Cubic model}

Let us consider cubic theory \eqref{eq:YLaction}. We assume that the theory has already been renormalized and fine-tuned to criticality within a specific scheme. We now place this critical theory on the manifold $S^1_\beta\times\mathbb{R}^{d-1}$. Since the background is locally flat, the previously fixed renormalization suffices to cancel ultraviolet divergences, allowing us to focus on finite-temperature corrections. The leading contribution to the one-point function $\braket{\phi}$ arises from the one-loop tadpole diagram:
\begin{equation}
    \vcenter{\hbox{\begin{tikzpicture}[baseline=(current bounding box.center)]
        \draw[thick] (0,-0.5) -- (0,0);
        \draw[thick] (0,0.5) circle (0.5);
        \fill (0,-0.5) circle (2pt);
    \end{tikzpicture}}}
    = -i g G(p = 0)  \left( \frac{1}{\beta}\sum_n \int \frac{d^{d-1} k}{(2\pi)^{d-1}} \frac{1}{\omega_n^2 + k^2 } - \int \frac{d^d k}{(2\pi)^d} \frac{1}{k^2}\right) = \infty\,.
\end{equation}

As seen in the expression above, the calculation is ill-defined. While the term in the brackets (representing the thermal loop correction after vacuum subtraction) is finite, it is multiplied by the zero-momentum propagator $G(p=0)$. Since the theory is critical (massless), $G(0)$ diverges. It is straightforward to verify that higher orders of perturbation theory exhibit increasingly severe infrared divergences. These singularities indicate the breakdown of naive perturbation theory and necessitate a resummation of the infrared divergences.

Let us perform this resummation explicitly for the cubic $O(N)$ model \cite{Fei:2014yja,Fei:2014xta} on $S^1_\beta\times\mathbb{R}^{d-1}$ with mass term to resolve the issue with the infrared divergences. The action is:
\begin{equation}
    S_{\text{cubic}}=\int d\tau d^{d-1}x\left(\frac{1}{2}\left(\partial_{\mu}\phi_i\right)^2+\frac{1}{2}\left(\partial_{\mu}\sigma\right)^2+\frac12 m^2 \sigma^2+\frac{g_1\sigma\phi_i^2}{2}+\frac{g_2\sigma^3}{3!}\right)\,.
\end{equation}

\begin{figure}[]
\centering
\begin{tikzpicture} 
\draw[color=black] (0,0) to (0,1); 
\draw[color=black] (0,1.2) circle [radius=0.2]; 
\node at (0,-0.5) {$\mathcal{T}^{}_{1}$};
\draw (0,1) node[vertex, inner sep=1pt]{};
\draw (0,0) node[vertex, inner sep=1pt]{};
\end{tikzpicture} 
\qquad
\begin{tikzpicture} 
\draw[color=black] (0,0) to (0,0.5); 
\draw[color=black] (0,0.5) to (0.3,1); 
\draw[color=black] (0,0.5) to (-0.3,1); 
\draw[color=black] (-0.3,1.2) circle [radius=0.2]; 
\draw[color=black] (0.3,1.2) circle [radius=0.2]; 
\node at (0,-0.5) {$\mathcal{T}_{3}$};
\draw (-0.3,1)node[vertex, inner sep=1pt]{};
\draw (0.3,1)node[vertex, inner sep=1pt]{};
\draw (0,0)node[vertex, inner sep=1pt]{} ;
\draw (0,0.5)node[vertex, inner sep=1pt]{};
\end{tikzpicture} 
\qquad
\begin{tikzpicture} 
\draw[color=black] (0,0) to (0,0.7);
\draw[color=black] (0,0.35) to (0.3,0.35); 
\draw[color=black] (0,0.7) to (0.3,1); 
\draw[color=black] (0,0.7) to (-0.3,1); 
\draw[color=black] (-0.3,1.2) circle [radius=0.2]; 
\draw[color=black] (0.3,1.2) circle [radius=0.2]; 
\draw[color=black] (0.5,0.35) circle [radius=0.2]; 
\node at (0,-0.5) {$\mathcal{T}_{5}$};
\draw (-0.3,1)node[vertex, inner sep=1pt]{};
\draw (0.3,1)node[vertex, inner sep=1pt]{};
\draw (0,0)node[vertex, inner sep=1pt]{};
\draw (0,0.35)node[vertex, inner sep=1pt]{};
\draw (0.3,0.35)node[vertex, inner sep=1pt]{};
\draw (0,0.7)node[vertex, inner sep=1pt]{};
\end{tikzpicture} 
\qquad
\begin{tikzpicture} 
\draw[color=black] (0,0) to (0,0.4);
\draw[color=black] (0,0.4) to (0.45,0.8); 
\draw[color=black] (0.45,0.8) to (0.25,1); 
\draw[color=black] (0.45,0.8) to (0.7,1); 
\draw[color=black] (0,0.4) to (-0.45,0.8); 
\draw[color=black] (-0.45,0.8) to (-0.25,1); 
\draw[color=black] (-0.45,0.8) to (-0.7,1); 
\draw[color=black] (0.25,1.2) circle [radius=0.2];
\draw[color=black] (0.7,1.2) circle [radius=0.2]; 
\draw[color=black] (-0.25,1.2) circle [radius=0.2];
\draw[color=black] (-0.7,1.2) circle [radius=0.2];
\node at (0,-0.5) {$\mathcal{T}^{(1)}_7$};
\draw (-0.25,1)node[vertex, inner sep=1pt]{};
\draw (-0.7,1)node[vertex, inner sep=1pt]{};
\draw (0.25,1)node[vertex, inner sep=1pt]{};
\draw (0.7,1)node[vertex, inner sep=1pt]{};
\draw (0.45,0.8)node[vertex, inner sep=1pt]{};
\draw (-0.45,0.8)node[vertex, inner sep=1pt]{};
\draw (0,0)node[vertex, inner sep=1pt]{};
\draw (0,0.4)node[vertex, inner sep=1pt]{};
\end{tikzpicture} 
\qquad
\begin{tikzpicture} 
\draw[color=black] (0,0) to (0,0.9);
\draw[color=black] (0,0.2) to (0.3,0.2); 
\draw[color=black] (0,0.7) to (0.3,0.7); 
\draw[color=black] (0,0.9) to (0.3,1.1); 
\draw[color=black] (0,0.9) to (-0.3,1.1); 
\draw[color=black] (-0.3,1.3) circle [radius=0.2]; 
\draw[color=black] (0.3,1.3) circle [radius=0.2]; 
\draw[color=black] (0.5,0.2) circle [radius=0.2];
\draw[color=black] (0.5,0.7) circle [radius=0.2];
\node at (0,-0.5) {$\mathcal{T}^{(2)}_7$};
\draw (-0.3,1.1)node[vertex, inner sep=1pt]{};
\draw (0.3,1.1)node[vertex, inner sep=1pt]{};
\draw (0,0)node[vertex, inner sep=1pt]{};
\draw (0,0.2)node[vertex, inner sep=1pt]{};
\draw (0.3,0.2)node[vertex, inner sep=1pt]{};
\draw (0,0.7)node[vertex, inner sep=1pt]{};
\draw (0.3,0.7)node[vertex, inner sep=1pt]{};
\draw (0,0.7)node[vertex, inner sep=1pt]{};
\draw (0,0.9)node[vertex, inner sep=1pt]{};
\end{tikzpicture} 
\caption{Leading most IR divergent contributions to the one-point function.}
\label{Tadpole}
\end{figure}
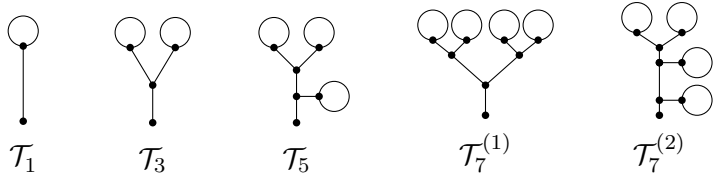
Because of the presence of $O(N)$ symmetry the field does not get condensed $\braket{\phi_i}=0$. Nonetheless, the field $\sigma$ has only $\mathcal{PT}$ symmetry that allows the existence of non-zero imaginary condensate of this field. To find this condensate $\braket{\sigma}$, we resum the most divergent tadpoles that contributes to it, which are presented in the Fig. \ref{Tadpole} up to the order\footnote{For brevity, we draw diagrams without distinguishing between the $\phi_i$ and $\sigma$ propagators.} $\mathcal{O}(g_1^{n_1}g_2^{n_2})$, $n_1+n_2=7$.
We note that the internal propagators are restricted to the $\sigma$ field, a direct consequence of the $O(N)$ symmetry.
We can note that each such diagram could be mapped to a bracket sequence or binary trees, and from that, we can immediately read off the combinatorial coefficients of these tadpoles in all orders \(
S^{\sigma}_k=\frac{C_k}{2^{2k+1}}=\frac{(2k)!}{2^{2k+1} k!(k+1)!}\), where $C_k$ are the Catalan numbers. These symmetry coefficients have been conjectured in \cite{Altherr:1991fu}. We can then resum all these IR divergences to obtain the leading-order term in the one-point function
\begin{equation}
\begin{aligned}\label{eq:vtadpolecubic}
   \langle \sigma \rangle&= v_0=-\sum_{k=0}^\infty S^\sigma_k\frac{(Ng_1 \Pi_0(0) +g_2 \Pi_0(m^2))^{k+1}g_2^k}{m^{2(2k+1)}}\\
   &=\frac{m^2}{g_2}\left(-1+\sqrt{1-\frac{g_2(Ng_1 \Pi_0(0) +g_2 \Pi_0(m^2) )}{m^4}}\right)\,,
\end{aligned}
\end{equation}
where we used that the expectation values $\braket{\phi_i^2}_0$ and $\braket{\sigma^2}_0$ in the free theory can be written in terms of $\Pi_0(m^2)$ (see Appendix \ref{App:FreeEnergyMassive}):
\begin{equation}\label{Pi0Definition}
    \Pi_0(m^2)\equiv\frac{1}{\beta}\sum_n \int \frac{d^{d-1} p}{(2\pi)^{d-1}} \frac{1}{p^2 + \omega_n^2 + m^2}, \quad \omega_n = \frac{2\pi n}{\beta}\,.
\end{equation}

In the conformal limit $m\rightarrow0$, each term in the sum individually diverges. However, after resummation, this limit is well-defined and finite. Consequently, we obtain
\begin{equation}\label{phi^3:vFromQuadraticEquation}
    v_0 = -i\sqrt{\frac{(N g_1+g_2)\Pi_0(0)}{g_2}} + \mathcal{O}(m^2) = -i\sqrt{\frac{\pi(N g_1 + g_2)}{180g_2}}T^2 + \mathcal{O}(m^2)\,,
\end{equation}
where we use the $d=6$ result $\Pi_0(0) = \frac{\pi T^4}{180}$.
Pure imaginary $v_0$ with negative imaginary part corresponds to pure imaginary couplings $g_1,g_2$ with positive imaginary parts. At finite temperature, fields $\phi_i$ and $\sigma$ have thermal masses $m_{\text{th},\phi}^2=g_1v_0>0$ and $m_{\text{th},\sigma}^2=g_2v_0>0$ in conformal limit. Note that for the $OSp(1|2)$ model with $N=-2$, $g_2=2g_1$ we have $v=0$ in agreement with the fact that one-point function is forbidden due to the supergroup symmetry \cite{Fei:2015kta,Klebanov:2021sos}. In $d=2$, Yang-Lee $N=0$ theory describes $M(2,5)$ minimal model \cite{Cardy:1985yy}, while the non-trivial $N=1$ fixed point describes $M(3,8)_D$ \cite{Fei:2014xta,Klebanov:2022syt,Katsevich:2024jgq}.

To generalize this procedure to other $\mathcal{PT}$-symmetric theories, we provide another approach to this problem. To see this, note that the binary trees in Fig. \eqref{Tadpole} are generated recursively: starting from a single line, one either splits it into two, each of which then generates its own binary tree giving \(v_0^2\), or terminates it by self-contraction, yielding \(\Pi_0(0)\):
\begin{equation}\label{eq:cubic1pteq}
    v_0=-\frac{g_2v_0^2}{2m^2}-\frac{Ng_1\Pi_0(0)+g_2\Pi_0(m^2)}{2m^2}\,.
\end{equation}
Quadratic equation on $v_0$ has two solutions:
\begin{equation}
    v_{0,\pm}=\frac{-m^2\pm\sqrt{m^4-g_2(Ng_1\Pi_0(0)+g_2\Pi_0(m^2))}}{g_2}\,,
\end{equation}
but only $v_{0,+}$ has correct small-$g$ expansion \eqref{eq:vtadpolecubic} and limit $v_0\rightarrow0$ as we send $m\rightarrow\infty$.

We are now in a position to compute the free energy of the cubic $O(N)$ model. To this end, we introduce an auxiliary source for the $\sigma$ field by adding the term $-h\sigma$ to the Lagrangian. This allows us to find that
\begin{equation}
    v_h = \frac{m^2}{g_2}\left(\sqrt{1 - \frac{2 g_2 h }{m^4} -  \frac{\pi(Ng_1+g_2)g_2T^4}{180m^4}} - 1\right),
\end{equation}
but on the other hand we can see that
\begin{equation}
    \frac{\partial}{\partial h} \left( \frac{F}{\beta V_{d-1}}\right) = v_h= \frac{m^2}{g_2}\left(\sqrt{1 - \frac{2 g_2 h }{m^4} -  \frac{\pi(Ng_1+g_2)g_2T^4}{180m^4}} - 1\right)\,.
\end{equation}
From this computation, we can easily compute the free energy of the theory in the conformal limit $m\rightarrow 0$:
\begin{equation}
    \frac{f_\text{cubic}}{T^d}=-\frac{(N+1)\Gamma(\frac{d}{2})\zeta(d)}{\pi^{\frac{d}{2}}}-\frac{1}{3g_2^2}\left(-\frac{\pi(Ng_1+g_2)g_2}{180}\right)^\frac{3}{2}+\mathcal{O}(g_1^{n_1}g_2^{n_2})\,, \label{eq:cubiconsimplemethod}
\end{equation}
where $n_1+n_2=2$.

\subsection{Quintic model}
\label{sec:quintic-trees}
Let us consider the quintic $O(N)$ model \cite{Gracey:2017okb,Klebanov:2021sos,Katsevich:2024jgq}:
\begin{equation}
    S_{\text{quintic}}=\int d\tau d^{d-1}x\left(\frac{1}{2}(\partial_\mu\phi)^2+\frac{1}{2}(\partial_\mu\sigma)^2+\frac{m^2\sigma^2}{2}+\frac{g_1\sigma(\phi_i^2)^2}{4!}+\frac{g_2\sigma^3\phi_i^2}{2 \cdot 3!}+\frac{g_3\sigma^5}{5!}\right)\,.
\end{equation}
Again, the leading contribution to the one-point function, $\braket{\sigma}=v_0$, appears after resumming IR divergent diagrams, and can be written in an equation analogous to \eqref{eq:cubic1pteq}:
\begin{equation}
    \begin{tikzpicture}[baseline={(c.base)}, line width=0.4pt]
  \node[inner sep=0pt] (c) at (0,0) {};
  \draw (-1.0,0) -- (0,0);
  \fill (0,0) circle (1.1pt);
  \node at (0.2,0.0) {\scriptsize \(v_0\)};
\end{tikzpicture}= -
\begin{tikzpicture}[baseline={(c.base)}, line width=0.4pt]
  \node[inner sep=0pt] (c) at (0,0) {};
  \draw (-1.0,0) -- (0,0);
  \fill (0,0) circle (1.1pt);
  \foreach \ang in {60,120,-60,-120}{
    \draw (0,0) -- ++(\ang:0.75);
    \fill ++(\ang:0.75) circle (1.1pt);
  }
  \node at (0.4,0.8) {\scriptsize \(v_0\)};
  \node at (0.4,-0.8) {\scriptsize \(v_0\)};
  \node at (-0.4,0.8) {\scriptsize \(v_0\)};
  \node at (-0.4,-0.8) {\scriptsize \(v_0\)};
\end{tikzpicture}-\begin{tikzpicture}[baseline={(c.base)}, line width=0.4pt]
  \node[inner sep=0pt] (c) at (0,0) {};
  \draw (-1.0,0) -- (0,0);
  \fill (0,0) circle (1.1pt);
  \draw (0,0) -- (0.0,0.85);
  \draw (0,0) -- (0.0,-0.85);
  \fill (0.0,0.85) circle (1.1pt);
  \fill (0.0,-0.85) circle (1.1pt);
  \node at (0.2,0.85) {\scriptsize \(v_0\)};
  \node at (0.2,-0.85) {\scriptsize \(v_0\)};
  \draw (0,0) .. controls (0.85,0.55) and (0.85,-0.55) .. (0,0);
  \node at (1.05,0) {\scriptsize \(\Pi_0(0)\)};
\end{tikzpicture}-
\begin{tikzpicture}[baseline={(c.base)}, line width=0.4pt]
  \node[inner sep=0pt] (c) at (0,0) {};
  \draw (-1.0,0) -- (0,0);
  \fill (0,0) circle (1.1pt);
  \draw (0,0) .. controls (0.55,0.85) and (-0.55,0.85) .. (0,0);
  \draw (0,0) .. controls (0.55,-0.85) and (-0.55,-0.85) .. (0,0);
  \node at (0,-0.9) {\scriptsize \(\Pi_0(0)\)};
  \node at (0,0.9) {\scriptsize \(\Pi_0(0)\)};
\end{tikzpicture}\,.
\end{equation}
Corresponding equation on $v_0$:
\begin{equation} 
\begin{aligned}
v_0 &= -\frac{g_3v_0^4}{4!m^2}-\frac{(Ng_2\Pi_0(0)+g_3\Pi_0(m^2))v_0^2}{4m^2}\\
&-\frac{N(N+2)g_1\Pi_0^2(0)+6Ng_2\Pi_0(0)\Pi_0(m^2)+3g_3\Pi^2_0(m^2)}{24m^2}\,,
\end{aligned}
\end{equation}
where in $d=\frac{10}{3}$ we have $\Pi_0(0)=\frac{3\Gamma(\frac{5}{3})\zeta(\frac{4}{3})T^{\frac{4}{3}}}{4\pi^{\frac{5}{3}}}$ (see Appendix \ref{App:FreeEnergyMassive}). This biquadratic equation has four solutions, and only one solution has a correct small $g$ expansion and limit $v_0\rightarrow0$ for $m\rightarrow\infty$. This one solution at the conformal limit:
\begin{equation}
    v_0=-i\sqrt{\frac{3\Pi_0(0)(Ng_2+g_3)}{g_3}\left(1-\sqrt{1-\frac{(N(N+2)g_1+6Ng_2+3g_3)g_3}{9(Ng_2+g_3)^2}}\right)}+\mathcal{O}(m^2)\,.
\end{equation}
Again, this solution for $v_0$ corresponds to pure imaginary $g_1,g_2,g_3$ with positive imaginary part. For $N=0$, this solution will take a simple form which doesn't depend on couplings:
\begin{gather}\label{phi^5:vFromQuarticEquation}
   v_0=-i\sqrt{(3-\sqrt{6})\Pi_0(0)}\,.
\end{gather}
Note that for the $OSp(1|4)$ model with $N=-4$, $g_2=\frac{2}{3}g_1$, and $g_3=\frac{8}{3}g_1$, we have $v_0=0$, in agreement with the absence of the one-point function $\braket{\sigma}$ due to supergroup symmetry \cite{Klebanov:2021sos}. It was conjectured that the two-dimensional conformal quintic model at $N=0$ describes the $M(2,7)$ minimal model \cite{Katsevich:2025ojk}, while the $N=1$ fixed points describe $M(5,14)_D$ and $M(5,16)_D$ \cite{Katsevich:2024jgq}.

\section{Thermal Ordered Operators}\label{sec:thermord}
In this section, we will introduce the notion of thermal ordered operators, which will allow us to completely avoid the infrared divergences and obtain a well-defined perturbative expansion. 
\subsection{Cubic model}
We have observed that naive perturbation theory in flat spacetime is plagued by IR divergences at finite temperatures. As indicated by the diagrams discussed previously, the primary origin of these divergences is the existence of tadpole sub-diagrams. To cure this pathology, we can redefine our operators to explicitly account for these contributions, a procedure that we will call as ``thermal normal ordering''. Consider the cubic Yang-Lee model:
\begin{equation}
    S_{\text{YL}}=\int d\tau d^{d-1}x\left(\frac12 (\partial_\mu \phi)^2 + \frac{g}{3!} \phi^3\right)\,,
\end{equation}
where to leading order the bare coupling is simply replaced by the renormalized one. We shift the field $\phi$ by a constant imaginary background $ v$, leading to the following transformation of the action:
\begin{equation}
    S_\text{YL}=\int d\tau d^{d-1}x\left(\frac12 (\partial_\mu \phi)^2+\frac{g}{3!}\phi^3 + \frac{gv}{2}\phi^2+\frac{gv^2}{2}\phi+ \frac{gv^3}{3!}\right)\,.
\end{equation}
Let us introduce thermally ordered operators \cite{Coleman:1974bu}: 
\begin{equation}
\begin{aligned}
    &\phi=:\phi:_T,\quad\phi^2=:\phi^2:_T+\Pi_0(m_\text{th}^2)\,,\\
    &\phi^3=:\phi^3:_T+3\Pi_0(m_\text{th}^2):\phi:_T\,,
\end{aligned}
\end{equation}
where $\Pi_0(m^2_{th})$ is given by \eqref{Pi0Definition}, and $m_\text{th}$ is thermal mass generated after performing the ``thermal normal ordering". After that we get the following action
\begin{equation}
\begin{aligned}
    S_\text{YL}&=\int d\tau d^{d-1}x\left(\frac12 (\partial_\mu\phi)^2+\frac{m^2_{\rm th}}{2}\phi^2+\frac{g}{3!}:\phi^3:_T+\frac{gv-m^2_{\text{th}}}{2}:\phi^2:_T\right.\\
    &\left.+\frac{g(v^2+\Pi_0(m_\text{th}^2))}{2}:\phi:_T +\frac{gv-m^2_{\text{th}}}{2}\Pi_0(m_\text{th}^2)+\frac{g v^3}{3!}\right)\,.
\end{aligned}
\end{equation}
Although the theory remains formally equivalent to the original model for any choice of $v$ and $m_{\rm th}$, we specifically select these parameters such that the interaction terms $:\phi:_T$ and $:\phi^2:_T$ are absent in the perturbation theory. This requirement imposes the following self-consistency (gap) equation:
\begin{equation}
    v^2 = - \Pi_0(m_\text{th}^2),\quad m^2_{\rm th} = g v\,.
\end{equation}
Solving the gap equation perturbatively in the coupling constant $g$ yields two distinct solutions for $v$. We resolve this ambiguity by determining $v$ at leading order and fixing the sign choices through comparison with \eqref{phi^3:vFromQuadraticEquation}, which gives:
\begin{equation}
    v=-i\sqrt{\Pi_0(m^2_\text{th})},\quad m_{\text{th}}^2=gv>0\,.
\end{equation}
Note that again this $v$ corresponds to pure imaginary $g$ with positive imaginary part. We reproduce thermal mass $m_\text{th}^2=gv$ from \cite{Altherr:1991fu}.

Consequently, the leading contribution to the free energy in the $\epsilon$-expansion is given by:
\begin{equation}
    f_{\text{YL}}=f_\text{free}(m^2_{\rm th})+\frac{gv-m^2_{\text{th}}}{2}\Pi_0(m_\text{th}^2)+\frac{gv^3}{3!}+\mathcal{O}(g^2)\,.
\end{equation}
It is worth noting that the gap equation derived above corresponds precisely to the stationarity condition of $F$ with respect to the parameters $v$ and $m_\text{th}^2$:
\begin{equation}
\begin{aligned}
    &\frac{\partial  f_{\rm YL}}{\partial v}=\frac{g}{2}(\Pi_0(m_\text{th}^2)+v^2)= 0\,,\\
    &\frac{\partial f_{\rm YL}}{\partial m_\text{th}^2}=\frac{1}{2}\Pi'_0(m_\text{th}^2)(gv- m_\text{th}^2) = 0\,.
\end{aligned}
\end{equation}
Using gap equations, we can obtain
\begin{equation}
    f_{\text{YL}}=f_\text{free}(gv)-\frac{m_\text{th}^2}{6}\Pi_0(m_\text{th}^2)+\mathcal{O}(g^2)=f_\text{free}(gv)+\frac{gv^3}{3!}+\mathcal{O}(g^2)\,.
\end{equation}
Note that once we take into account loop corrections, the free energy will be subject to the renormalization that will cancel UV divergences in the higher order of perturbation theory, as we will check in Section \ref{sec:renormfree}. We will study this function and the loop corrections to free energy in subsequent sections.

\subsection{Quintic model}
Let us show that the previous technique could be applied to other theories. For that, let us study the quintic Tricritical Yang-Lee model:
\begin{equation}
    S_{TYL}=\int d\tau d^{d-1}x\left(\frac12 (\partial_\mu\phi)^2 + \frac{g}{5!}\phi^5\right)\,.
\end{equation}
Again, we introduce shift in the imaginary direction $\phi \to \phi +  v$:
\begin{equation}
     S_{TYL}=\int d\tau d^{d-1}x\left(\frac12 (\partial_\mu\phi)^2  +  \frac{g}{5!} \phi^5
     + \frac{g v}{24}\phi^4 + \frac{g v^2}{12} \phi^3
     +  \frac{g v^3}{12}\phi^2
     + \frac{ g v^4}{24} \phi + \frac{g v^5}{5!}\right)\,.
\end{equation}
In contrast to the cubic model discussed above, the present theory receives tadpole contributions to the one-point functions of both $\phi$ and $\phi^2$.
After introducing ``thermal normal ordering'' the action becomes
\begin{equation}\label{eq:quiticactionordered}
\begin{aligned}
    S_{TYL}&=\int d\tau d^{d-1}x\left(\frac12 (\partial_\mu\phi)^2+\frac{m_\text{th}^2}{2}\phi^2 +\frac{g}{4}\left(\frac{1}{2}\Pi^2_0(m_\text{th}^2)+v^2\Pi_0(m_\text{th}^2) + \frac{v^4}{6}\right):\phi:_T\right.\\
    &+\frac{1}{2}\left(\frac{gv}{2}\Pi_0(m_\text{th}^2) + \frac{g v^3}{6}-m_\text{th}^2\right):\phi^2:_T+\frac{g(v^2+\Pi_0(m_\text{th}^2))}{12}:\phi^3:_T \\
    &\left.+ \frac{gv}{24}:\phi^4:_T +\frac{g}{5!} :\phi^5:_T+\frac{gv}{8}\Pi_0^2(m_\text{th}^2)+\frac{gv^3-6m_\text{th}^2}{12}\Pi_0(m_\text{th}^2)+ \frac{g v^5}{5!}\right)\,.
\end{aligned}
\end{equation}
Demanding that coupling constants in front of  $:\phi:_T$ and $:\phi^2:_T$ cancel, we can find that
\begin{equation}\label{phi^5:GapEquationMass}
    \frac{1}{2}\Pi^2_0(m_\text{th}^2)+ v^2\Pi_0(m_\text{th}^2) + \frac{v^4}{6}=0\,, \quad m_\text{th}^2=\frac{g v}{2}\Pi_0(m_\text{th}^2) + \frac{g v^3}{6}\,.
\end{equation}
Solving biquadratic equation on $v$, we obtain
\begin{equation}
    v^2=-( 3 \pm \sqrt{6})\Pi_0(m_{\text{th}}^2)\,,
\end{equation}
and substituting this back into the expression for the thermal mass \eqref{phi^5:GapEquationMass}, we find
\begin{equation}
    m_{\text{th}}^2=\frac{g v^3}{6}\left(-2 \pm \sqrt{6} \right) >0\,. 
\end{equation}
By solving the gap equation perturbatively in the coupling constant $g$, we obtain four distinct solutions for $v$. We resolve this ambiguity by first determining $v$ at leading order and then fixing the sign choices by comparison with \eqref{phi^5:vFromQuarticEquation} leading to:
\begin{equation}
    v=-i\sqrt{(3-\sqrt{6})\Pi_0(m^2_\text{th})},\quad m_{\text{th}}^2=-\frac{g v^3}{6}\left(2 + \sqrt{6} \right) >0\,.
\end{equation}
Note that again this $v$ corresponds to pure imaginary $g$ with positive imaginary part. So $:\phi^4:_T$ coefficient in \eqref{eq:quiticactionordered} $\frac{gv}{24}>0$ and $m_\text{th}^2>0$. It then follows that the free energy to the leading order in $g$ is given by
\begin{equation}
    f_{\rm TYL}= f_\text{free}(m_\text{th}^2)+\frac{gv}{8}\Pi_0^2(m_\text{th}^2)+\frac{gv^3-6m_\text{th}^2}{12}\Pi_0(m_\text{th}^2)+\frac{g v^5}{5!}\,.
\end{equation}
Note that, once again, the gap equations can be obtained by minimizing the thermal free energy
\begin{equation}
\begin{aligned}
    &\frac{\partial  f_{\rm TYL}}{\partial v}=\frac{g}{4}\left(\frac{1}{2} \Pi_0^2(m_\text{th}^2) + v^2 \Pi_0(m_\text{th}^2)+\frac{v^4}{6}\right) = 0\,,\\
    &\frac{\partial f_{\rm TYL}}{\partial m_\text{th}^2}=\frac{1}{2}\Pi'_0(m_\text{th}^2) \left(\frac{gv}{2} \Pi_0(m_\text{th}^2) + \frac{g v^3}{6}- m_\text{th}^2\right) = 0\,.
\end{aligned}
\end{equation}
Using gap equations, we can obtain the final form of the free energy to the leading order in $g$:
\begin{equation}
    f_{\rm TYL}=f_\text{free}(m_\text{th}^2)-\frac{3m_\text{th}^2}{10}\Pi_0(m_\text{th}^2)=f_\text{free}(m_{\text{th}}^2)-g v^5\frac{12+5 \sqrt{6}}{60} \,.
\end{equation}

\section{Cubic $O(N)$ model}\label{sec:ThermalO(N)}
In this section, we only consider the cubic $O(N)$ model in $\mathbb{S}^1 \times \mathbb{R}^{5-\epsilon}$ and compute the leading and subleading corrections to the free energy in the $\epsilon$-expansion. Again, we use the thermal normal ordering that will allow us to have a consistent and reliable perturbation theory.

\subsection{Gap equation}
\label{sec:renormfree}
Let us consider massless cubic $O(N)$ model in $\mathbb{S}^1\times\mathbb{R}^{5-\epsilon}$ \cite{Fei:2014yja,Fei:2014xta}:
\begin{equation}\label{eq:actionO(N)}
    S_{\text{cubic}}=\int d\tau d^{d-1}x\left(\frac{1}{2}\left(\partial_{\mu}\phi_{i,0}\right)^2+\frac{1}{2}\left(\partial_{\mu}\sigma_0\right)^2+\frac{g_{1,0}\sigma_0\phi_{i,0}^2}{2}+\frac{g_{2,0}\sigma_0^3}{3!}\right)\,.
\end{equation}
Let us note that $\mathcal{PT}$ symmetry acts only on the field $\sigma_0$: $\sigma_0 \to -\sigma_0, i \to -i$. $\mathbb{Z}_2$ symmetry acts on fields $\phi_{i,0}$: $\phi_{i,0}\rightarrow-\phi_{i,0}$. This field $\sigma_0$ can get condensed. In contrast, fields $\phi_{i,0}$ can not be condensed because of the $O(N)$ symmetry. 

To define the theory in a controlled way, we introduce a regularization scheme. Throughout this section, all computations are performed in the bare theory, and the renormalization procedure is applied only at the end. As a consequence, the thermal ordering prescription introduced in the previous section may generate divergences in the shifted theory. For instance, the solution of the gap equation might be divergent. However, these divergences cancel once the correlators are expressed in the original unshifted theory.

To implement thermal ordering, it is sufficient to shift the field $\sigma_0$ by a constant imaginary background $v$ as $\sigma_0 \rightarrow \hat\sigma_0+v$ , where we choose $v$ to cancel all tadpole diagrams in the definition of thermally ordered operators. Under this shift, the action (\ref{eq:actionO(N)}) becomes
\begin{equation}
    \begin{aligned}
    S_\text{cubic}&=\int d\tau d^{d-1}x\left(\frac12 (\partial_\mu\phi_{i,0})^2+\frac12 (\partial_\mu\hat\sigma_0)^2+\frac{g_{1,0}\hat\sigma_0\phi_{i,0}^2}{2}+\frac{g_{2,0}\hat\sigma_0^3}{3!}+\frac{g_{1,0}v}{2}\phi_{i,0}^2+\right.\\
    &\left.+\frac{g_{2,0}v}{2}\hat\sigma_0^2+\frac{g_{2,0}v^2}{2}\hat\sigma_0+\frac{g_{2,0}v^3}{3!}\right)\,.
    \end{aligned}
\end{equation}
Using that, the thermally ordered operators have the following form
\begin{equation}
\begin{aligned}
&\phi_{i,0}=:\phi_{i,0}:_T,\quad\hat\sigma_0=:\hat\sigma_0:_T,\quad\phi_{i,0}^2=:\phi_{i,0}^2:_T+N\Pi_0(m_{\text{th},\phi,0}^2)\,,\\
&\hat\sigma_0^2=:\hat\sigma_0^2:_T+\Pi_0(m_{\text{th},\sigma,0}^2),\quad\hat\sigma_0^3=:\hat\sigma_0^3:_T+3\Pi_0(m_{\text{th},\sigma,0}^2):\hat\sigma_0:_T\,,
\end{aligned}
\end{equation}
we obtain
\begin{equation}
\begin{aligned}
    &S_\text{cubic}=\int d\tau d^{d-1}x\left(\frac{1}{2}\left(\partial_{\mu}\phi_{i,0}\right)^2+\frac{m^2_{\text{th},\phi,0}}{2}\phi_{i,0}^2+\frac{1}{2}\left(\partial_{\mu}\hat\sigma_0\right)^2
    +\frac{m^2_{\text{th},\sigma,0}}{2}\hat\sigma_0^2\right.\\ 
    &+\frac{g_{1,0}:\hat\sigma_0\,\phi_{i,0}^2:_T}{2}+\frac{g_{2,0}:\hat\sigma^3_0:_T}{3!}+\frac{g_{1,0}v-m_{\text{th},\phi,0}^2}{2}:\phi_{i,0}^2:_T+\frac{g_{2,0}v-m_{\text{th},\sigma,0}^2}{2}:\hat\sigma_0^2:_T\\
    &+\frac{1}{2}\left(Ng_{1,0}\Pi_0(m_{\text{th},\phi,0}^2)+g_{2,0}\Pi_0(m_{\text{th},\sigma,0}^2)+g_{2,0}v^2\right):\hat\sigma_0:_T\\
    &\left.+\frac{N(g_{1,0}v-m_{\text{th},\phi,0}^2)}{2}\Pi_0(m^2_{\text{th},\phi,0})+\frac{g_{2,0}v-m_{\text{th},\sigma,0}^2}{2}\Pi_0(m^2_{\text{th},\sigma,0})+\frac{g_{2,0} v^3}{3!}\right)\,.
\end{aligned}
\end{equation}
Demanding that coupling constants in front of  $:\phi_{i,0}^2:_T$, $:\hat\sigma_0^2:_T$ and $:\hat\sigma_0:_T$ cancel, we can obtain gap equation on $v$ and thermal masses $m^2_{\text{th},\phi,0}$, $m^2_{\text{th},\sigma,0}$:
\begin{equation}\label{GapEquationO(N)}
Ng_{1,0}\Pi_0\left(g_{1,0}v\right)+g_{2,0}\Pi_0\left(g_{2,0}v\right)+g_{2,0}v^2=0\,,\quad m_{\text{th},\phi,0}^2=g_{1,0}v\,,\quad m_{\text{th},\sigma,0}^2=g_{2,0}v\,.
\end{equation}
Using the gap equations, the action can be written as
\begin{equation}
\begin{aligned}
    S_\text{cubic}&=\int d\tau d^{d-1}x\left(\frac{1}{2}(\partial_{\mu}\phi_{i,0})^2+\frac{g_{1,0}v}{2}\phi_{i,0}^2+\frac{1}{2}\left(\partial_{\mu}\hat\sigma_0\right)^2+\frac{g_{2,0}v}{2}\hat\sigma_0^2\right.\\
    &\left.+\frac{g_{1,0}:\hat\sigma_0\,\phi_{i,0}^2:_T}{2}+\frac{g_{2,0}:\hat\sigma^3_0:_T}{3!}+\frac{g_{2,0} v^3}{3!}\right)\,.
\end{aligned}
\end{equation}
Propagators of the fields $\phi_{i,0}$ and $\hat{\sigma}_0$ read as 
\begin{equation}\label{O(N):BarePropagators}
    \widetilde{G}_{\phi,0}(\omega_n,p)=\frac{1}{\omega_n^2 + p^2 + g_{1,0} v}\,, \quad \widetilde{G}_{\hat{\sigma},0 }(\omega_n,p)=\frac{1}{\omega_n^2 + p^2 + g_{2,0} v}
\end{equation}
with bosonic Matsubara frequencies $\omega_n = \tfrac{2\pi n}{\beta}$.

\begin{figure}[t]
\centering

\begin{subfigure}[b]{0.28\textwidth}
\centering
\begin{tikzpicture}
\draw[color=black,densely dashed] (0,0) -- (0,1);
\draw[color=black] (0,1.3) circle [radius=0.3];
\draw[color=black,densely dashed] (-0.3,1.3) -- (0.3,1.3);
\draw (-0.3,1.3)node[vertex, inner sep=1pt]{};
\draw (0.3,1.3)node[vertex, inner sep=1pt]{};
\node at (0,-0.45) {$\mathcal{T}^{(1)}_{3}$};
\draw (0,1)node[vertex, inner sep=1pt]{};
\draw (0,0)node[vertex]{};
\end{tikzpicture}
\caption{$\mathcal{O}(g_{1,0}^3)$ contribution}
\label{fig:Tadpole_1}
\end{subfigure}
\begin{subfigure}[b]{0.28\textwidth}
\centering
\begin{tikzpicture}
\draw[color=black,densely dashed] (0,0) -- (0,1);

\draw[color=black] (0,1.3) ++(180:0.3) arc (180:0:0.3);

\draw[color=black,densely dashed] (0,1.3) ++(0:0.3) arc (0:-180:0.3);

\draw[color=black] (-0.3,1.3) -- (0.3,1.3);
\draw (-0.3,1.3) node[vertex, inner sep=1pt]{};
\draw (0.3,1.3) node[vertex, inner sep=1pt]{};

\node at (0,-0.45) {$\mathcal{T}^{(2)}_{3}$};

\draw (0,1) node[vertex, inner sep=1pt]{};
\draw (0,0) node[vertex]{};
\end{tikzpicture}

\caption{$\mathcal{O}(g_{1,0}^2g_{2,0})$ contribution}
\label{fig:Tadpole_2}
\end{subfigure}
\begin{subfigure}[b]{0.28\textwidth}
\centering
\begin{tikzpicture}
\draw[color=black,densely dashed] (0,0) -- (0,1);
\draw[color=black,densely dashed] (0,1.3) circle [radius=0.3];
\draw[color=black,densely dashed] (-0.3,1.3) -- (0.3,1.3);
\draw (-0.3,1.3)node[vertex, inner sep=1pt]{};
\draw (0.3,1.3)node[vertex, inner sep=1pt]{};
\node at (0,-0.45) {$\mathcal{T}^{(3)}_{3}$};
\draw (0,1)node[vertex, inner sep=1pt]{};
\draw (0,0)node[vertex]{};
\end{tikzpicture}
\caption{$\mathcal{O}(g_{2,0}^3)$ contribution}
\label{fig:Tadpole_3}
\end{subfigure}

\caption{Lowest-order contributions to the one-point function.}
\label{fig:tadpolesO(N)}
\end{figure}
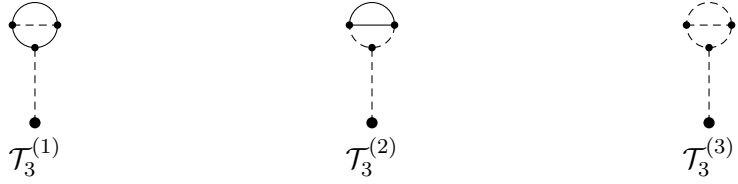

\subsection{One-point function}
Let us note that the actual value that is finite and well-defined would be the quantity  $\braket{\sigma} = Z^{-\frac{1}{2}}_\sigma (\braket{\hat\sigma_0} + v )$, where $Z_\sigma$ is the wave function renormalization of field $\sigma$, that can be computed by studying the renormalization of the theory in flat spacetime. Thus, we expect that the solution to the gap equation \eqref{GapEquationO(N)} could contain divergences, but after taking into account UV divergences coming from loop computations of the $\braket{\hat{\sigma}_0}$ and $Z_\sigma$, all these divergences should cancel and we get a finite answer. The renormalization of couplings was performed up to five loops \cite{Fei:2014yja,Fei:2014xta,Gracey:2015tta,Kompaniets:2021hwg}, here we use only one-loop results:
\begin{equation}\label{RenormaizationO(N)}
    \begin{aligned}
    &g_{1,0}=\mu^{\frac{\epsilon}{2}}\left(g_1+\frac{(N-8)g^3_{1}-12 g^2_1g_2+g_1g^2_2}{12(4\pi)^3 \epsilon}\right)\,,\\ 
    &g_{2,0}=\mu^{\frac{\epsilon}{2}}\left(g_2-\frac{4Ng^3_1-Ng^2_1g_2+3g^3_2}{4(4\pi)^3\epsilon}\right)\,, \\
    &Z_{\phi}=1-\frac{g^2_{1}}{3 (4 \pi )^3 \epsilon },\quad Z_{\sigma}=1-\frac{Ng^2_{1}+g_2^2}{6 (4\pi)^3 \epsilon}\,.
    \end{aligned}
\end{equation}
Let us follow this procedure. Firstly, we solve the gap equation \eqref{GapEquationO(N)} (see Appendix ~\ref{App:solvevac}) and indeed obtain a UV divergence
\begin{equation}\label{uvdiv1}
        \frac{v}{T^{2-\frac{\epsilon}{2}}}= \frac{i}{\epsilon}\frac{10 Ng_1^3 +(N g_1 +6 g_2) \left(N g_1^2 +g_2^2\right)}{4608 \sqrt{5} \pi ^{5/2}  \sqrt{Ng_1 g_2+g^2_2}}
        -i\sqrt{ \frac{\pi}{180}\frac{N g_{1}+g_{2}}{g_{2}}} + \dots,
\end{equation}
where from \eqref{eq:vacsol} we have extracted the UV-divergent pole and the leading finite term that is necessary for the subsequent cancellation of UV divergences. This solution for $v$ corresponds to purely imaginary couplings $g_1$ and $g_2$ with positive imaginary parts.
As we have explained before, this pole must be canceled by the one-point function of the field $\braket{\hat{\sigma}_0}$, which also contains UV divergences from higher loops (for instance, see Figure~\ref{fig:tadpolesO(N)}). Thus, at two loops the one-point function of $\hat{\sigma}_0$ receives the following contributions:
\begin{equation}
    \langle \hat{\sigma}_0 \rangle=-\frac{Ng_{1,0}^3}{2}\,\mathcal{T}^{(1)}_3-\frac{Ng_{1,0}^2 g_{2,0}}{4}\,\mathcal{T}^{(2)}_3-\frac{g_{2,0}^3}{4}\,\mathcal{T}^{(3)}_3,
\end{equation}
where each diagram is drawn in the Figure~\ref{fig:tadpolesO(N)}.
\begin{equation}
\begin{aligned}
    \mathcal{T}_3^{(1)}&=\frac{T^2}{g_{2,0}v}\sum \limits_{n_p,n_q}\int\frac{d^{d-1}pd^{d-1}q}{(2\pi)^{2(d-1)}}\frac{1}{(P^2+g_{1,0}v)^2(Q^2+g_{1,0}v)((P+Q)^2+g_{2,0}v)}\,,\\
    \mathcal{T}_3^{(2)}&=\frac{T^2}{g_{2,0}v}\sum \limits_{n_p,n_q}\int\frac{d^{d-1}pd^{d-1}q}{(2\pi)^{2(d-1)}}\frac{1}{(P^2+g_{2,0}v)^2(Q^2+g_{1,0}v)((P+Q)^2+g_{1,0}v)}\,,\\
    \mathcal{T}_3^{(3)}&=\frac{T^2}{g_{2,0}v}\sum \limits_{n_p,n_q}\int\frac{d^{d-1}pd^{d-1}q}{(2\pi)^{2(d-1)}}\frac{1}{(P^2+g_{2,0}v)^2(Q^2+g_{2,0}v)((P+Q)^2+g_{2,0}v)}\,.
\end{aligned}
\end{equation}
These diagrams can be obtained as a derivative with respect to mass of sunset diagram $\mathcal{I}_2(m_1^2,m_2^2,m_3^2)$, which was computed in Appendix \ref{App:I_2} (see \eqref{I2FinalEpsExpansion}):
\begin{equation}
\begin{aligned}
    &\mathcal{T}_3^{(1)}=-\frac{1}{g_{2,0}v}\frac{\partial}{\partial m_1^2}\mathcal{I}_2(g_{1,0}v,g_{1,0}v,g_{2,0}v)=\frac{T^{2d-8}}{g_{2,0}v}\left(\frac{1}{6912\pi^2\epsilon}+b-\frac{\sqrt{g_{1,0}v}}{768\pi^3T}+\mathcal{O}(g_{1,0}^{n_1}g_{2,0}^{n_2})\right)\,,\\
    &\mathcal{T}_3^{(2)}=-\frac{1}{g_{2,0}v}\frac{\partial}{\partial m_1^2}\mathcal{I}_2(g_{2,0}v,g_{1,0}v,g_{1,0}v)=\frac{T^{2d-8}}{g_{2,0}v}\left(\frac{1}{6912\pi^2\epsilon}+b-\frac{\sqrt{g_{2,0}v}}{768\pi^3T}+\mathcal{O}(g_{1,0}^{n_1}g_{2,0}^{n_2})\right)\,,\\
    &\mathcal{T}_3^{(3)}=-\frac{1}{g_{2,0}v}\frac{\partial}{\partial m_1^2}\mathcal{I}_2(g_{2,0}v,g_{2,0}v,g_{2,0}v)=\frac{T^{2d-8}}{g_{2,0}v}\left(\frac{1}{6912\pi^2\epsilon}+b-\frac{\sqrt{g_{2,0}v}}{768\pi^3T}+\mathcal{O}(g_{1,0}^{n_1}g_{2,0}^{n_2})\right)\,
\end{aligned}\label{uvdiv2}
\end{equation}
with $n_1+n_2=1$.

Now we can combine the \eqref{uvdiv1} and \eqref{uvdiv2} and see that after wave function renormalization all divergences cancel and we obtain a finite answer $\braket{\sigma}=Z_\sigma^{-\frac{1}{2}}(\langle\hat\sigma_0\rangle+v)$:
\begin{equation}\label{eq:1ptfunctioneps}
\begin{aligned}
    &\frac{\braket{\sigma}^{}}{\mu^{-\gamma_\sigma}T^{\Delta_\sigma}}=-\frac{i}{6}\sqrt{\frac{\pi(Ng_1+g_2)}{5g_2}}\left(1+\epsilon\frac{A}{12}\right)+\frac{Ng_{1}^2+g_{2}^2}{96\pi g_{2}}\\
    &+\left(\frac{Ng_1+g_2}{5g_2}\right)^{\frac{1}{4}}\frac{N\left(-i g_{1}\right)^{\frac{5}{2}}+\left(-i g_{2}\right)^{\frac{5}{2}}}{48\sqrt{6}\pi^{\frac{7}{4}}g_{2}}+i\sqrt{\frac{5g_2}{Ng_1+g_2}}\times\\
    &\times\Bigg(\frac{2N g_1^3+Ng^2_1g_2 +g_2^3}{27648 \pi ^{5/2} g_2}(A-41472\pi^2b)-\frac{(N g_1+g_2)\left(N g_1^3+g_2^3\right)}{7680 \pi ^{5/2} g_2^2}\log(4 \pi  e^{-\gamma_E })\\
   &+ \frac{ \left(Ng_{1}^2+g_2^2\right)^2}{3072 \pi ^{5/2} g_2^2}+\frac{N g_1  \left(6 Ng_1^3-(N+4) g_1^2 g_2-6 g_1 g_2^2+5 g_2^3\right)}{23040 \pi ^{5/2} g_2^2 }\log \left(\frac{\mu}{T}\right)\Bigg)\\
   &+i\left(\frac{5g_2}{Ng_1+g_2}\right)^{\frac{1}{4}}\frac{N(-ig_{1})^{\frac{5}{2}}(Ng_{1}^2-2g_{1}g_{2}+g_{2}^2)}{512\sqrt{6}\pi^{\frac{13}{4}}g_{2}^2}+\mathcal{O}(\epsilon^{n_1}g_1^{n_2}g_2^{n_3})|_{2n_1+n_2+n_3=3}\,,
\end{aligned}
\end{equation}
where $A$ is defined in \eqref{Adef}, and some $\log\frac{\mu}{T}$ were absorbed in the anomalous dimension of the $\sigma$ field
\begin{equation}
    \Delta_\sigma=\frac{d-2}{2}+\gamma_\sigma,\quad\gamma_\sigma=\frac{Ng_{1}^2+g_{2}^2}{12(4\pi)^3}+\mathcal{O}(g^{n_1}_1g^{n_2}_2)|_{n_1+n_2=4}\,.
\end{equation}
Let us stress that substituting fixed point $g_{1,\star}$ and $g_{2,\star}$ into $\frac{\braket{\sigma}}{\mu^{-\gamma_\sigma}T^{\Delta_\sigma}}$ yields a $\mu$-independent result.
To compare this one-point function with the large \(N\) result \(\langle \bar{\sigma}_\star \rangle\) and with the two-dimensional one-point functions extracted from the OPE coefficient \eqref{eq:OPE1pt}, we first introduce the normalization factor \(\mathcal{N}_\sigma\) through the two-point function of \(\sigma\) in the massless cubic \(O(N)\) model on \(\mathbb{R}^d\) at \(T=0\),
\begin{equation}
    \braket{\sigma(x)\sigma(y)}=\frac{\mathcal{N}_\sigma^2}{\mu^{2\gamma_\sigma}|x-y|^{2\Delta_\sigma}}\,.
\end{equation}
The corresponding diagrams were computed in \cite{Giombi:2024zrt}, from which one finds, to leading order in the couplings,
\begin{equation}
    \mathcal{N}_\sigma^2
    =\frac{1}{4\pi^3}
    -\frac{\epsilon\bigl(1-\log(\pi e^{\gamma_E})\bigr)}{8\pi^3}
    -\frac{(N g_{1}^2+g_{2}^2)\bigl(5+3\log(\pi e^{\gamma_E})\bigr)}{9216\pi^6}\,.
\end{equation}
With this normalization, we define
\begin{gather}\label{BarSigma}
  \bar{\sigma}=\frac{\sigma}{\mathcal{N}_\sigma \mu^{-\gamma_\sigma}}\,,
\end{gather}
so that $\langle \bar{\sigma}(x)\bar{\sigma}(y)\rangle=\tfrac{1}{|x-y|^{2\Delta_\sigma}}$.

\subsection{Two-point functions and thermal masses}
In this subsection, we analyze the two-point functions to derive the thermal masses for the $\sigma$ and $\phi_i$ fields. Since the connected part of the two-point function remains invariant under field shifts, we can equivalently perform the renormalization within the shifted theory to extract the thermal masses.

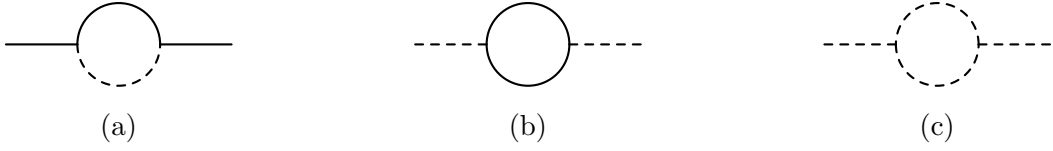
\begin{figure}[t]
\centering
\begin{minipage}{0.3\textwidth}
\centering
\begin{tikzpicture}[scale=2.5, line cap=round, line join=round]
    \draw[line width=0.8pt] (-0.6,0) -- (-0.22,0);
    \draw[line width=0.8pt] (0.22,0) -- (0.6,0);

    \draw[line width=0.8pt] (0.22,0) arc[start angle=0,end angle=180,radius=0.22];
    \draw[line width=0.8pt,dashed] (-0.22,0) arc[start angle=180,end angle=360,radius=0.22];
\end{tikzpicture}

\vspace{0.3em}
(a)
\end{minipage}
\hfill
\begin{minipage}{0.3\textwidth}
\centering
\begin{tikzpicture}[scale=2.5, line cap=round, line join=round]
    \draw[line width=0.8pt,dashed] (-0.6,0) -- (-0.22,0);
    \draw[line width=0.8pt,dashed] (0.22,0) -- (0.6,0);

    \draw[line width=0.8pt] (0.22,0) arc[start angle=0,end angle=180,radius=0.22];
    \draw[line width=0.8pt] (-0.22,0) arc[start angle=180,end angle=360,radius=0.22];
\end{tikzpicture}

\vspace{0.3em}
(b)
\end{minipage}
\hfill
\begin{minipage}{0.3\textwidth}
\centering
\begin{tikzpicture}[scale=2.5, line cap=round, line join=round]
    \draw[line width=0.8pt,dashed] (-0.6,0) -- (-0.22,0);
    \draw[line width=0.8pt,dashed] (0.22,0) -- (0.6,0);
    \draw[line width=0.8pt,dashed] (0.22,0) arc[start angle=0,end angle=180,radius=0.22];
    \draw[line width=0.8pt,dashed] (-0.22,0) arc[start angle=180,end angle=360,radius=0.22];
\end{tikzpicture}

\vspace{0.3em}
(c)
\end{minipage}
\caption{Diagram (a) contributes to leading order in $\Sigma_{\phi,T}(\omega_n,p)$, while diagrams (b) and (c) contribute to $\Sigma_{\sigma,T}(\omega_n,p)$.}
\label{fig:Sigma_Self_Energies}
\end{figure}

Performing the renormalization at finite temperature, we obtain the following expressions:
\begin{equation}\label{ExactФФProp}
    \begin{aligned}
    &Z_\phi^{-1}\widetilde{G}^{-1}_{\phi}(\omega_n,p)=\omega_n^2 + p^2 + g_{1,0} \left(v+\braket{\hat{\sigma}_0}\right)+\Sigma_{\phi,T}(\omega_n,p)\,,\\
    &Z_\sigma^{-1}\widetilde{G}^{-1}_{\sigma}(\omega_n,p)=\omega_n^2 + p^2 + g_{2,0} \left(v+\braket{\hat{\sigma}_0}\right)+\Sigma_{\sigma,T}(\omega_n,p)\,.
\end{aligned}
\end{equation}
The leading-order contributions to $\Sigma_{\phi,T}(\omega_n,p)$ and $\Sigma_{\sigma,T}(\omega_n,p)$ are shown in figure~\ref{fig:Sigma_Self_Energies} and are given by 
\begin{equation}\label{Sigma2}
    \begin{aligned}
    &\Sigma^{(2)}_{\phi,T}(\omega_n,p)=-g^2_{1,0}K_{T}\left(g_{1,0}v,g_{2,0}v;\omega_n,p\right)\,,\\
    &\Sigma^{(2)}_{\sigma,T}(\omega_n,p)=-\frac{Ng^2_{1,0}}{2}K_{T}\left(g_{1,0}v,g_{1,0}v;\omega_n,p\right) -\frac{g^2_{2,0}}{2}K_{T}\left(g_{2,0}v,g_{2,0}v;\omega_n,p\right) \,, 
\end{aligned}
\end{equation}
where 
\begin{equation}
   K_{T}\left(m^2_1,m^2_2;p_0,p\right)= \frac{1}{\beta}\sum \limits_{n_q}\int\frac{d^{d-1}q}{(2\pi)^{d-1}}\frac{1}{ (q^2_0+q^2+m^2_1)((q_0+p_0)^2+(q+p)^2+m^2_2)}\,,
\end{equation}
and its properties are discussed in appendix~\ref{APP:KTIntegral}.
Using the results of the previous subsection, we arrive at
\begin{equation}
    \begin{aligned}
    &Z_{\phi} g_{1,0} \left(v+\braket{\hat{\sigma}_0}\right)=\bigg[Z_{\phi} g_{1,0} \left(v+\braket{\hat{\sigma}_0}\right)\bigg]_{\rm fin}+i\sqrt{\frac{\pi(Ng_1+g_2)}{180g_2}}\frac{g^2_{1}\left(g_{1}+g_2\right)}{(4\pi)^3 \epsilon}\,,\\
    &Z_{\sigma} g_{2,0} \left(v+\braket{\hat{\sigma}_0}\right)=\bigg[Z_{\sigma} g_{2,0} \left(v+\braket{\hat{\sigma}_0}\right)\bigg]_{\rm fin}+i\sqrt{\frac{\pi(Ng_1+g_2)}{180g_2}}\frac{Ng^2_{1}+g^3_2}{(4\pi)^3 \epsilon}\,.
\end{aligned}
\end{equation}
Utilizing \eqref{KTDivPart}, we observe that all divergences in \eqref{ExactФФProp} cancel at cubic order in the couplings. Consequently, the connected two-point functions are finite, allowing us to define the thermal masses through the following conditions:
\begin{equation}
     \widetilde{G}^{-1}_{\phi}(\omega_n=0,p^2=-M^2_{\phi,\rm th.})=0\,, \quad \quad  \widetilde{G}^{-1}_{\sigma}(\omega_n=0,p^2=-M^2_{\sigma,\rm th.})=0\,.
\end{equation}
Or equivalently:
\begin{equation}\label{EquationOnThermalMass}
    \begin{aligned}
    &M^2_{\rm th,\phi}=\bigg[Z_{\phi} g_{1,0} \left(v+\braket{\hat{\sigma}_0}\right)\bigg]_{\rm fin}+\bigg[Z_{\phi}\Sigma_{\phi,T}(0,iM_{\rm th,\phi})\bigg]_{\rm fin}\,,\\
    &M^2_{\rm th,\sigma}=\bigg[Z_{\sigma} g_{2,0} \left(v+\braket{\hat{\sigma}_0}\right)\bigg]_{\rm fin}+\bigg[Z_{\sigma}\Sigma_{\sigma,T}(0,iM_{\rm th,\sigma})\bigg]_{\rm fin}\,.
\end{aligned}
\end{equation}
Solving these equations perturbatively, we obtain the following expressions for the thermal masses at the fixed point:
\begin{equation}\label{eq:themalmass}
    \begin{aligned}
    &\beta^2M^2_{\rm th,\phi}=-ig_{1}  \sqrt{\frac{N g_{1}+g_{2}}{g_{2}}}\sqrt{\frac{\pi}{180}}+\frac{g_{1}(Ng_{1}^2 -2 g_{1} g_{2}+g_{2}^2)}{96 \pi g_{2}}+\mathcal{O}(g_1^{n_1}g_2^{n_2})\,,\\
    &\beta^2M^2_{\rm th,\sigma}=-ig_{2}  \sqrt{\frac{N g_{1}+g_{2}}{g_{2}}}\sqrt{\frac{\pi }{180}} +\mathcal{O}\left(g_1^{n_1}g_2^{n_2}\right)\,,
\end{aligned}
\end{equation}
where $n_1+n_2=\frac{5}{2}$.

\subsection{Thermal free energy}
In this subsection, we compute the thermal free energy in the $\epsilon$ expansion. Up to two loops, the free energy density admits the expansion shown in figure~\ref{Fig:FreeEnergyDiagramsO(N)}:
\begin{equation}
    f_{\text{cubic}}= 
    Nf_{\text{free}}(g_{1,0}v)+f_{\text{free}}(g_{2,0}  v)+\frac{g_{2,0}v^3}{3!}  -\frac{Ng^2_{1,0}}{4}\mathcal{I}^{(1)}_2-\frac{g^2_{2,0}}{12}\mathcal{I}^{(2)}_2+\mathcal{O}(g_{1,0}^{n_1}g_{2,0}^{n_2})\,,
\end{equation}
where $n_1+n_2=4$ and 
\begin{equation}
    \begin{aligned}
        \mathcal{I}^{(1)}_2&=\frac{1}{\beta^2}\sum \limits_{n_p,n_q}\int\frac{d^{d-1}pd^{d-1}q}{(2\pi)^{2(d-1)}}\frac{1}{(P^2+g_{1,0}v)(Q^2 + g_{1,0}v)((P+Q)^2+g_{2,0} v)}\,,\\
        \mathcal{I}^{(2)}_2&=\frac{1}{\beta^2}\sum \limits_{n_p,n_q}\int\frac{d^{d-1}pd^{d-1}q}{(2\pi)^{2(d-1)}}\frac{1}{(P^2+g_{2,0} v)(Q^2 + g_{2,0}v)((P+Q)^2 + g_{2,0} v)}\,.
    \end{aligned}
\end{equation}
Here we introduced $P=(p_0,p)$ and $Q=(q_0,q)$, where $p_0=\frac{2\pi n_p}{\beta}$ and $q_0=\frac{2\pi n_q}{\beta}$. Since the masses depend on the bare couplings $g_{i,0}$, evaluation of these contributions is cumbersome. Instead, we compute the corresponding diagrams by performing a small-mass expansion up to cubic order in thermal masses (see appendix \ref{App:I_2} for details). Combining the result with \eqref{eq:FreeEnergyMassive}, we obtain
\begin{equation}\label{freeEnergyExpansionPert}
\begin{aligned}
    \frac{f_\text{cubic}}{T^{6-\epsilon}}&= (N+1)\pi^3\left(-\frac{2}{945}+\epsilon\frac{3-2 \log\left(\pi e^{\gamma_E-2\frac{\zeta'(6)}{\zeta(6)}}\right)}{1890}\right)+v\frac{ N g_{1,0}+g_{2,0}}{T^2}\frac{\pi}{360}\left(1+\frac{A}{6}\epsilon\right)\\
    &-v^2\frac{Ng_{1,0}^2+g_{2,0}^2}{192 \pi  T^4}+v^{\frac{5}{2}}\frac{Ng_{1,0}^{\frac{5}{2}}+g_{2,0}^{\frac{5}{2}}}{120 \pi ^2 T^5}+v^3\frac{Ng_{1,0}^3+g_{2,0}^3}{384\pi^3T^6}\left(\frac{1}{\epsilon}-\frac{\log(4\pi e^{-\gamma_E})}{2}\right)+\frac{g_{2,0}v^3}{6 T^{6-\epsilon}}
    \\
    &-\frac{Ng^2_{1,0}}{4 T^d}\mathcal{I}^{(1)}_2-\frac{g^2_{2,0}}{12 T^d}\mathcal{I}^{(2)}_2+\mathcal{O}(g_{1,0}^{n_1}g_{2,0}^{n_2})\,,
\end{aligned}
\end{equation}
where $A$ is defined in \eqref{Adef}, and $n_1+n_2=4$. Note that we have UV divergent terms that are coming from the expansion of the free energy of free massive theory around $d=6$, along with that, we have UV divergences in the diagrams, but after renormalization of bare coupling constants through renormalized ones \eqref{RenormaizationO(N)} we expect that all these divergences should cancel. Thus, using \eqref{I2FinalEpsExpansion}, we have
\begin{figure}[t]
\centering
\begin{tikzpicture}
  \draw[color=black] (0.6,0) arc[start angle=0, end angle=180, radius=0.6];

  \draw[color=black,densely dashed] (-0.6,0)
    arc[start angle=180, end angle=360, radius=0.6];

  \draw (-0.6,0) node[vertex]{};
  \draw (0.6,0) node[vertex]{};
  \draw[color=black] (-0.6,0) -- (0.6,0);

  \node at (0,-1) {$\mathcal{I}^{(1)}_2$};
\end{tikzpicture}
\qquad
\qquad
\begin{tikzpicture} 
\draw[color=black, densely dashed]  (0,0) circle [radius=0.6]; 
\draw(-0.6,0)node[vertex]{} ;
\draw (0.6,0)node[vertex]{} ;
\draw[color=black, densely dashed] (-0.6,0)to (0.6,0);
\node at (0,-1) {$\mathcal{I}^{(2)}_2$};
\end{tikzpicture} 
\caption{Contributions to the free energy in the cubic $O(N)$ model at two-loop order.}
\label{Fig:FreeEnergyDiagramsO(N)}
\end{figure}
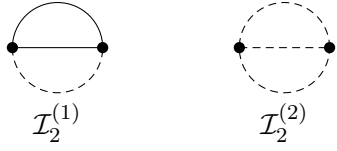
\begin{equation}
\begin{aligned}
     -\frac{Ng^2_{1,0}}{4}\mathcal{I}^{(1)}_2-\frac{g^2_{2,0}}{12}\mathcal{I}^{(2)}_2& =  \frac{2 N g_{1,0}^3 +Ng_{1,0}^2 g_{2,0} +g_{2,0}^3}{27648\pi^2 } vT^{2d-8}\left(\frac{1}{\epsilon}+6912 \pi^2 b\right) \\
    &-\frac{ T^{2 d-8}}{4608 \pi ^3}  \left(2 Ng_{1,0}^2\left(g_{1,0} v\right)^{\frac{3}{2}} +Ng_{1,0}^2 \left(g_{2,0} v\right)^{\frac{3}{2}} +g_{2,0}^2\left(g_{2,0} v\right)^{\frac{3}{2}}\right)\,,
\end{aligned}
\end{equation}
where $b=4.68376\cdot10^{-5}$ is given in \eqref{b_numericalvalue}.  
Putting everything together, and using renormalization of bare couplings to cubic order in coupling constants, we will get that the divergence will cancel out, and we get the following result in $d=6-\epsilon$:
\begin{equation}\label{freeEnergyExpansionRenorm}
    \begin{aligned}
    &\frac{f_\text{cubic}}{T^{6-\epsilon}}=(N+1)\pi^3\left(-\frac{2}{945}+\epsilon\frac{3-2\log\left(\pi e^{\gamma_E-2\frac{\zeta'(6)}{\zeta(6)}}\right)}{1890}\right)\\
    &-i\left(\frac{Ng_1+g_2}{5g_2}\right)^{\frac{1}{2}}\frac{Ng_1+g_2}{3240}\pi^{\frac{3}{2}}\left(1+\epsilon\frac{A+\log\left(\frac{\mu^2}{T^{2}}\right)}{4}\right)+\frac{Ng_1+g_2}{5g_2}\frac{Ng_{1}^{2}+g_{2}^{2}}{6912}
    \\
    &+\left(\frac{Ng_1+g_2}{5g_2}\right)^{\frac{5}{4}} 
    \frac{N(-ig_{1})^{\frac{5}{2}}+(-i g_{2})^{\frac{5}{2}}}{4320\sqrt{6}\pi^{\frac{3}{4}}}-i\left(\frac{Ng_1+g_2}{5g_2}\right)^{\frac{3}{2}}\frac{N g_{1}^{3}+g_{2}^{3}}{165888\pi^{\frac{3}{2}}}\log{\left(4\pi e^{-\gamma_E}\frac{T^2}{\mu^2}\right)}\\
    &+i\left(\frac{Ng_1+g_2}{5g_2}\right)^{\frac{1}{2}}\Bigg(\frac{2 Ng_1^{3}+Ng_1^{2}g_2+g_2^3}{995328\pi^{\frac{3}{2}}}\left(A-3\log\left(\frac{\mu^2}{T^{
    2}}\right)\right)+\frac{(Ng_{1}^{2} +g_{2}^{2})^2}{110592\pi^{\frac{3}{2}}g_{2}}\\
    &-\frac{ \sqrt{\pi}(2Ng_{1}^{3}+Ng_{1}^{2}g_{2} +g_{2}^{3})}{24}b\Bigg)+i\left(\frac{Ng_1+g_2}{5g_2}\right)^{\frac{3}{4}}\frac{N(-i g_{1})^{\frac{5}{2}}(Ng_{1}^{2}-2 g_{1}g_{2}+g_{2}^{2})}{27648\sqrt{6} \pi^{\frac{9}{4}} g_{2}}\\
    &+\mathcal{O}(\epsilon ^{n_1}g^{n_2}_1g^{n_3}_{2})|_{2n_1+n_2+n_3=4}\,.
    \end{aligned}
\end{equation}
Let us stress that, as in the case of one-point function, substituting fixed point $g_{1,\star}$ and $g_{2,\star}$ into $f_\text{cubic}$ yields a $\mu$-independent result.

\section{Results for the large $N$, Yang-Lee, and $N=1$ cubic models}\label{sec:results}
In this section, we provide numerical results for the thermal mass, one-point function, and thermal free energy for different fixed points (large $N$, $N=0$ Yang-Lee, and $N=1$) at various integer dimensions.

\subsection{Cubic large $N$ model}\label{sec:cubiclargeN}
We begin with the cubic \(O(N)\) model at large \(N\) and compare its thermal free energy with the large \(N\) analysis of Section \ref{sec:LargeN}. The \(6-\epsilon\) expansion of the cubic theory is known up to five loops \cite{Fei:2014yja,Fei:2014xta,Gracey:2015tta,Kompaniets:2021hwg}. 
The large \(N\) fixed point in \(d=6-\epsilon\) has real couplings and $\mathcal{PT}$ symmetry is broken. This indicates a thermal instability, as first observed in \cite{Altherr:1991fu} for the $N=0$ model with real coupling. This conclusion is consistent with the large $N$ analysis of section~\ref{sec:LargeN}. In contrast, in $d=6+\epsilon$ we have imaginary couplings and $\mathcal{PT}$ symmetry. Thus, for convenience,  we work in \(d=6+\epsilon\) with purely imaginary couplings:
\begin{equation}
     \begin{aligned}
     g_{1,\star}&=i\sqrt{\frac{6\epsilon(4\pi)^3}{N}}\left(1+\frac{22}{N}+\frac{726}{N^2}+\left(\frac{155}{6N}+\frac{1705}{N^2}\right)\epsilon+\mathcal{O}\left(\frac{1}{N^3},\frac{\epsilon^2}{N}\right)\right)\,,\\
     g_{2,\star}&=6i\sqrt{\frac{6\epsilon(4\pi)^3}{N}}\left(1+\frac{162}{N}+\frac{68766}{N^2}+\left(\frac{215}{2N}+\frac{86335}{N^2}\right)\epsilon+\mathcal{O}\left(\frac{1}{N^3},\frac{\epsilon^2}{N}\right)\right)\,.
\end{aligned}
\end{equation}
The thermal masses \eqref{eq:themalmass} in the large $N$ fixed point in $d=6+\epsilon$ dimensions are given by
\begin{equation}
\begin{aligned}
    \frac{M^2_{\text{th},\phi}}{\pi^2T^2}&=\frac{4\sqrt{\epsilon}}{3\sqrt{5}}-\frac{2\epsilon}{3}+\frac{8\epsilon^{\frac{5}{4}}}{3\sqrt{3}5^{\frac{1}{4}}}-\frac{(1-12\gamma_E+1440\zeta'(-3))\epsilon^{\frac{3}{2}}}{18\sqrt{5}}-\frac{2\;5^{\frac{1}{4}}\epsilon^{\frac{7}{4}}}{\sqrt{3}}\\
    &-\frac{12}{N}(\sqrt{5\epsilon}-4\epsilon)-\frac{12}{N^2}(1141\sqrt{5\epsilon}-2416\epsilon)+\mathcal{O}\left(\epsilon^2,\frac{\epsilon^{\frac{5}{4}}}{N},\frac{\sqrt{\epsilon}}{N^3}\right)\,,\\
    \frac{M_{\text{th},\sigma}^2}{\pi^2T^2}&=\frac{8\sqrt{\epsilon}}{5}\left(1+\frac{95}{N}+\frac{65975}{2N^2}\right)+\mathcal{O}\left(\epsilon^{\frac{5}{4}},\frac{\sqrt{\epsilon}}{N^3}\right)\,.
\end{aligned}
\end{equation}
Note that the contribution of $\Sigma^{(2)}_{\phi,T}$ in \eqref{Sigma2} is of order $g_1^2$ and therefore enters only at order $N^{-1}$. As a result, the perturbative solution for $M_{\text{th},\phi}$ from \eqref{EquationOnThermalMass} at leading order in $N$ can be consistently extended up to order $\epsilon^{\frac{7}{4}}$. Then the leading \(\mathcal{O}(1)\) term in \(M^2_{\text{th},\phi}\) coincides with the large \(N\) result \eqref{eq:largeNmass}, providing a nontrivial check for the results in section \ref{sec:ThermalO(N)}.

Normalized one-point function $\braket{\bar{\sigma}}$ \eqref{BarSigma} in the large $N$ model in $d=6+\epsilon$:
\begin{equation}
\begin{aligned}
    &\frac{\braket{\bar{\sigma}}_{\text{large }N}}{\pi^2\sqrt{N}T^{\Delta_\sigma}}=-\frac{i}{3\sqrt{30}}+\frac{i\sqrt{\epsilon}}{6\sqrt{6}}-\frac{\sqrt{2}i\epsilon^{\frac{3}{4}}}{9\;5^{\frac{1}{4}}}+\frac{(17-12\gamma_E+1440\zeta'(-3))i\epsilon}{72\sqrt{30}}+\frac{5^{\frac{1}{4}}i\epsilon^{\frac{5}{4}}}{6\sqrt{2}}\\
    &+\frac{4.077i}{N}(1-1.3683\sqrt{\epsilon}+1.3483\epsilon^{\frac{3}{4}}+2.8603\epsilon-7.2393\epsilon^{\frac{5}{4}})\\
    &+\frac{1516.86i}{N^2}(1-1.8787\sqrt{\epsilon}+2.7485\epsilon^{\frac{3}{4}}+2.0938\epsilon-8.5259\epsilon^{\frac{5}{4}})+\mathcal{O}\left(\epsilon^{\frac{3}{2}},\frac{1}{N^3}\right)\,.
\end{aligned}
\end{equation}
The leading \(\mathcal{O}(\sqrt{N})\) term in \(\braket{\bar{\sigma}}_{\text{large }N}\) coincides with the large \(N\) result \eqref{eq:largeN1pt}, providing a nontrivial check for the results in section \ref{sec:ThermalO(N)}.

Thermal free energy (\ref{freeEnergyExpansionRenorm}) in the large $N$ fixed point in $d=6+\epsilon$:
\begin{equation}
\begin{aligned}
    &\frac{f_{\text{large }N}}{N\pi^3T^{6+\epsilon}}=-\frac{2}{945}+\frac{\sqrt{\epsilon}}{405\sqrt{5}}-\frac{13-4\log\left(\pi e^{\gamma_E-2\frac{\zeta'(6)}{\zeta(6)}}\right)}{3780} \epsilon+\frac{4\epsilon^{\frac{5}{4}}}{675\sqrt{3}5^{\frac{1}{4}}}\\
    &-\frac{1-4\log(4\pi e^{2 \gamma_E })+1440\zeta'(-3)}{3240\sqrt{5}} \epsilon^{\frac{3}{2}}-\frac{\epsilon^{\frac{7}{4}}}{27 \sqrt{3}5^{\frac{3}{4}}}\\
    &-\frac{2.1164\cdot10^{-3}}{N}\left(1+20.348\sqrt{\epsilon}-47.374\epsilon+26.290\epsilon^{\frac{5}{4}}+169.66\epsilon^{\frac{3}{2}}-421.53\epsilon^{\frac{7}{4}}\right)\\
    &-\frac{27.719}{N^2}\left(\sqrt{\epsilon}-2.8380\epsilon+3.2556\epsilon^{\frac{5}{4}}+5.7544\epsilon^{\frac{3}{2}}-17.621\epsilon^{\frac{7}{4}}\right)+\mathcal{O}\left(\epsilon^2,\frac{\sqrt{\epsilon}}{N^3}\right)\,.
\end{aligned}
\end{equation}
Again, the leading \(\mathcal{O}(N)\) term in \(f_{\text{large }N}\) coincides with the large \(N\) result \eqref{eq:largeNON}, providing another nontrivial check for the results in section \ref{sec:ThermalO(N)}.
\subsection{$N=0$ Yang-Lee model}
Let us move to the $N=0$ Yang-Lee cubic model \cite{Fisher:1978pf}. The $6-\epsilon$ expansion for this theory was performed up to six loops \cite{Fisher:1978pf,deAlcantaraBonfim:1980pe,Gracey:2015tta,Borinsky:2021jdb,Schnetz:2025wtu}. The IR fixed point up to two loops is
\begin{equation}
\begin{aligned}\label{YLFixedPoint}
    g_{1,\star}=0,\quad g_{2,\star}=i\sqrt{\frac{2(4\pi)^3\epsilon}{3}}\left(1+\frac{125}{324}\epsilon+\mathcal{O}(\epsilon^2)\right)\,.
\end{aligned}
\end{equation}
In $d=2$, $N=0$ Yang-Lee model is described by $M(2,5)$ minimal model \cite{Cardy:1985yy}. To determine the exact thermal mass in $d=2$, we use \eqref{eq:Deltaeff}. The only relevant operator in the \(M(2,5)\) minimal model is \(\phi_{1,2}\sim\sigma\), and its OPE takes the form
\begin{equation}
    \sigma\times\sigma\sim 1+i\sigma\,.
\end{equation}
Since the theory is non-unitary, the true ground state is not the identity but the state \(\ket{\Omega}=\ket{\sigma}\), whose scaling dimension is $\Delta_\sigma=-\frac{2}{5}$. The lowest state that appears with a nonzero OPE coefficient in the \(\sigma\times\sigma\) OPE and lies above the ground state is the identity operator. It follows that \cite{Itzykson:1986pk}
\begin{equation}
    \Delta_{\text{eff},\sigma}=\Delta_{I}-\Delta_{\sigma}=\frac{2}{5} \quad \rightarrow  \quad m_{\rm th} \beta  =2\pi\Delta_{\text{eff},\sigma}=\frac{4\pi}{5}\,.
\end{equation}
Finally, substituting \eqref{YLFixedPoint} into  \eqref{eq:themalmass}, we obtain
\begin{equation}
\begin{aligned}
    \frac{M_{\text{th},\sigma}^2}{\pi^2T^2}=
    \begin{cases}
        \frac{16}{25},\quad\quad\quad\quad\quad\quad\; d=2\,,\\
        \frac{4}{3}\sqrt{\frac{2\epsilon}{15}}+\mathcal{O}(\epsilon^{\frac{5}{4}}),\quad d=6-\epsilon\,.\\
    \end{cases}
\end{aligned}
\end{equation}
Formally setting $\epsilon=4$ into the leading-order term yields a perturbative result that deviates from the exact value by $23.3\%$.

Next, we can compare the normalized one-point function in the Yang--Lee model in \(d=2\) with its \(d=6-\epsilon\) expansion. In two dimensions, \eqref{eq:1ptfunction2D} gives $ \braket{\bar{\sigma}}=(2\pi)^{\Delta_\sigma}C_{\sigma\sigma\sigma}\,T^{\Delta_\sigma}$, where the \(M(2,5)\) OPE coefficient is \cite{Dotsenko:1984nm}\footnote{We use normalization with opposite sign of $\sigma$ comparing to \cite{Dotsenko:1984nm}.}
\begin{equation}
    C_{\sigma\sigma\sigma}=-\frac{i}{5}\frac{\Gamma^{\frac{3}{2}}(\frac{1}{5})\Gamma^{\frac{1}{2}}(\frac{2}{5})}{\Gamma^{\frac{3}{2}}(\frac{4}{5})\Gamma^{\frac{1}{2}}(\frac{3}{5})}\approx-1.91131i\,.
\end{equation}
Substituting \eqref{YLFixedPoint} into \eqref{eq:1ptfunctioneps} and using \eqref{BarSigma}, we have
\begin{figure}[]
    \centering
        \centering
        \includegraphics{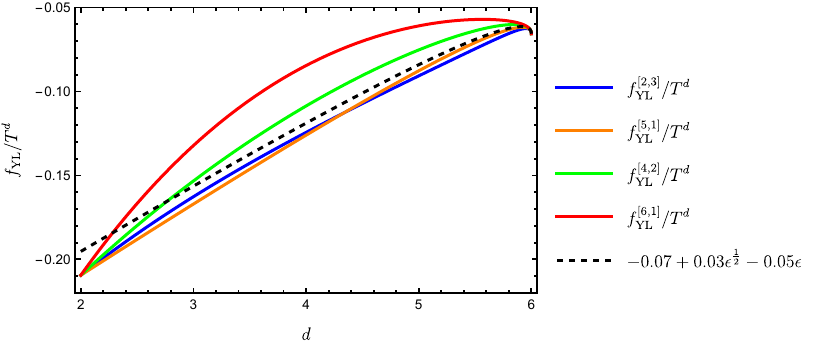}
  
    \caption{Two-sided Pad\'e extrapolations and $\mathcal{O}(\epsilon)$ expansion for the thermal free energy $\frac{f_\text{YL}(d)}{T^d}$  in the Yang-Lee theory.}
    \label{fig:YL}
\end{figure}
\begin{equation}\label{eq:oneptYL}
    \frac{\braket{\bar{\sigma}}_{\text{YL}}}{T^{\Delta_\sigma}}=\begin{cases}
        -0.9163i, & d=2\,,\\
        -\frac{\pi^2i}{3\sqrt{5}}+\frac{\pi^2i\sqrt{\epsilon}}{3\sqrt{6}}-\frac{2^{\frac{7}{4}}\pi^2i\epsilon^{\frac{3}{4}}}{9\, \cdot 15^{\frac{1}{4}}}+1.0900i\epsilon+\mathcal{O}(\epsilon^{\frac{3}{2}}), & d=6-\epsilon\,.
    \end{cases}
\end{equation}
Note that the $\epsilon^{\frac{5}{4}}$ term is absent in this case. Pad\'e extrapolation gives $\braket{\bar{\sigma}}_{\text{YL}}^{[1,2]}=-0.9116iT^{\Delta_\sigma}$ in $d=2$, which differs from an exact result by $0.5\%$. The Pad\'e extrapolation is performed in the variable \(t\equiv\epsilon^{1/4}\), in terms of which the expansion \eqref{eq:oneptYL} becomes a degree-$5$ polynomial in \(t\) with a vanishing \(t^5\) term.

Finally, utilizing the expansion \eqref{freeEnergyExpansionRenorm}, we determine the free energy as a function of $d$. In the $d=2$ limit, this result can be compared against the exact value derived from the effective central charge $c_{\text{eff}}=\frac{2}{5}$ using \eqref{eq:freeenergydensity}
\begin{equation}\label{eq:YLexp}
    \frac{f_{\text{YL}}}{T^{6-\epsilon}} = 
    \begin{cases} 
        -\frac{\pi}{15}, & d=2, \\[6pt]
        -0.0656 + 0.0280\sqrt{\epsilon} - 0.0464\epsilon\\
        + 0.0427\epsilon^\frac{5}{4}- 0.0101\epsilon^{\frac{3}{2}} + \mathcal{O}(\epsilon^2), & d=6-\epsilon\,.
    \end{cases}
\end{equation}
Surprisingly, the $\epsilon^{\frac{7}{4}}$ term is absent in this case and  the truncated expansion up to \(\mathcal{O}(\epsilon)\) yields
\(f^{\mathcal{O}(\epsilon)}_{\text{YL}}=-0.1953\,T^2\), which is within \(6.8\%\) of the exact \(M(2,5)\) result
\(f^{\text{exact}}_{M(2,5)}=-0.2094T^2\). This supports Cardy’s conjecture that \(M(2,5)\) admits a cubic Lagrangian description \cite{Cardy:1985yy}.
We also perform two-sided Pad\'e extrapolations for the Yang--Lee model imposing the \(d=2\) boundary condition (see Fig. \ref{fig:YL}). Unlike the sphere free energy case \cite{Giombi:2024zrt}, there exist two-sided Pad\'e approximants without poles in the range \(2<d<6\)\footnote{This is consistent with the relatively small magnitude of \(c_{\text{eff}}(2,5)=\frac{2}{5}\) in \(d=2\), compared to the central charge \(c(2,5)=-\frac{22}{5}\).}.
Table~\ref{tab:YL} summarizes various Pad\'e predictions for \(f_{\text{YL}}/T^d\) at \(d=3,4,5\).

It would be interesting to determine \(f_{\text{YL}}\) in \(d>2\) using independent nonperturbative methods beyond the \(\epsilon\)-expansion, such as the FRG \cite{An:2016lni,Zambelli:2016cbw,Rennecke:2022ohx,Benedetti:2026tpa}, high-temperature expansions \cite{Butera:2012tq}, the non-unitary bootstrap \cite{Gliozzi:2013ysa,Gliozzi:2014jsa,Hikami:2017hwv} (including finite-temperature implementations \cite{Iliesiu:2018fao,Barrat:2025wbi,Barrat:2025nvu}), and fuzzy sphere regularization \cite{ArguelloCruz:2025zuq,Fan:2025bhc,EliasMiro:2025msj}.

\begin{table}[]
    \centering
    \begin{tabular}{|c|c|c|c|c|}
    \hline
        Dimension & $d=5$ & $d=4$ & $d=3$ & $d=2$ \\\hline
        $f^{[2,3]}_{\text{YL}}/T^d$ & $-0.0907$ & $-0.1244$ & $-0.1627$ & $-0.2094$ \\\hline
        $f^{[5,1]}_{\text{YL}}/T^d$ & $-0.0877$ & $-0.1260$ & $-0.1670$ & $-0.2094$ \\\hline
        $f^{[4,2]}_{\text{YL}}/T^d$ & $-0.0752$ & $-0.1086$ & $-0.1533$ & $-0.2094$ \\\hline
        $f^{[6,1]}_{\text{YL}}/T^d$ & $-0.0609$ & $-0.0845$ & $-0.1322$ & $-0.2094$ \\\hline
        $f^{\mathcal{O}(\epsilon)}_{\text{YL}}/T^d$ & $-0.0841$ & $-0.1189$ & $-0.1564$ & $-0.1953$ \\\hline
    \end{tabular}
    \caption{Two-sided Pad\'e extrapolations and $\mathcal{O}(\epsilon)$ expansion for the thermal free energy $\frac{f_\text{YL}(d)}{T^d}$ in the Yang-Lee theory for $d=3,4,5$.}
    \label{tab:YL}
\end{table}

\subsection{Cubic $N=1$ model}
Let us consider the cubic $N=1$ model. The $6-\epsilon$ expansion for this theory is known up to five loops \cite{Fei:2014yja,Fei:2014xta,Gracey:2015tta,Kompaniets:2021hwg}. The non-trivial IR fixed point up to two loops is
\begin{equation}
\begin{aligned}\label{NEq1FixedPoint}
    g_{1,\star}&=40i\sqrt{\frac{6\pi^3\epsilon}{499}}\left(1+\frac{2633149}{7470030}\epsilon+\mathcal{O}(\epsilon^2)\right),\\
    g_{2,\star}&=48i\sqrt{\frac{6\pi^3\epsilon}{499}}\left(1+\frac{227905}{498002}\epsilon+\mathcal{O}(\epsilon^2)\right)\,.
\end{aligned}
\end{equation}
In $d=2$, the cubic $N=1$ model is conjectured to describe the $M(3,8)_D$ minimal model \cite{Fei:2014xta,Klebanov:2022syt,Katsevich:2024jgq}. To determine the exact thermal mass in two dimensions, we again use \eqref{eq:Deltaeff}. The OPEs of the relevant operators $\phi_{1,3}\sim\sigma$ and $\phi_{1,4}^-\sim\phi$ are
\begin{equation}
    \sigma\times\sigma\sim 1+i\sigma+...,\quad\sigma\times\phi\sim i\phi+...\,,
\end{equation}
\begin{figure}[]
    \centering
    \includegraphics{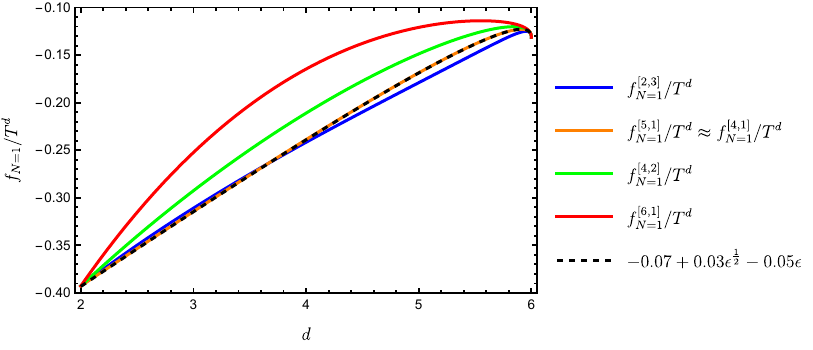}   
    \caption{Two-sided Pad\'e extrapolations and $\mathcal{O}(\epsilon)$ expansion for the thermal free energy $\frac{f_{N=1}(d)}{T^d}$ in the cubic $N=1$ model. Note that $f^{[4,1]}_{N=1}(d)$ and $f^{[5,1]}_{N=1}(d)$ differ from each other less than by $0.02\%$ in the range $2<d<6$, we denote it as $f^{[4,1]}_{N=1}(d)\approx f^{[5,1]}_{N=1}(d)$.}
    \label{fig:N=1}
\end{figure}
where $...$ stands for operators of higher dimension. The vacuum of the $M(3,8)_D$ model is the state $\ket{\Omega}=\ket{\sigma}$, whose dimension is $\Delta_\sigma=-\frac{1}{2}$. The lowest state appearing with a nonzero OPE coefficient in the \(\sigma\times\sigma\) OPE above the ground state is the identity operator, whereas the lowest state appearing in the \(\sigma\times\phi\) OPE is $\phi$ itself. It follows that
\begin{equation}
    \Delta_{\text{eff},\phi}=\Delta_\phi-\Delta_{\sigma}=\frac{5}{16}, \quad\Delta_{\text{eff},\sigma}=\Delta_I-\Delta_{\sigma}=\frac{1}{2}\,.
\end{equation}
Using \eqref{eq:Deltaeff} and the fixed point \eqref{NEq1FixedPoint} in \eqref{eq:themalmass}, we obtain the following expressions for the thermal masses of $\phi$ and $\sigma$:
\begin{subequations}\label{eq:Msigma}
\begin{align}
    \frac{M_{\text{th},\phi}^2}{\pi^2 T^2} &=
    \begin{cases}
        \dfrac{25}{64}, & d=2, \\[4pt]
        \dfrac{4}{3}\sqrt{\dfrac{55\epsilon}{499}} - \dfrac{10\pi^2\epsilon}{1497}
        + \mathcal{O}(\epsilon^{5/4}), & d=6-\epsilon,
    \end{cases}
    \\
    \frac{M_{\text{th},\sigma}^2}{\pi^2 T^2} &=
    \begin{cases}
        1, & d=2, \\[4pt]
        8\sqrt{\dfrac{11\epsilon}{2495}} + \mathcal{O}(\epsilon^{5/4}), & d=6-\epsilon.
    \end{cases}
\end{align}
\end{subequations}
Setting $\epsilon=4$, we find values that differ from the exact results for $M_{\text{th},\phi}$ and $M_{\text{th},\sigma}$ by $23.3\%$ and $1.3\%$, respectively.

In $d=2$, normalized one-point function \eqref{eq:1ptfunction2D} $\braket{\bar{\sigma}}=(2\pi)^{\Delta_\sigma}C_{\sigma\sigma\sigma}T^{\Delta_\sigma}$, where the $M(3,8)$ OPE coefficient \cite{Dotsenko:1984nm} is\footnote{We use normalization with opposite sign of $\sigma$ comparing to \cite{Dotsenko:1984nm}.}
\begin{equation}
    C_{\sigma\sigma\sigma}=-\frac{i}{2\sqrt[4]{8}}\frac{\Gamma(\frac{1}{8})\Gamma^{\frac{1}{2}}(\frac{5}{4})}{\Gamma(\frac{7}{8})\Gamma^{\frac{1}{2}}(\frac{3}{4})}\approx-1.76787i\,.   
\end{equation}
Substituting \eqref{NEq1FixedPoint} into \eqref{eq:1ptfunctioneps} and using \eqref{BarSigma}, we have
\begin{equation}
    \frac{\braket{\bar{\sigma}}_{N=1}}{T^{\Delta_\sigma}}=\begin{cases}
        -0.7053i, & d=2\,,\\
        -1.9921i + 1.8338i\sqrt{\epsilon} - 2.5776i\epsilon^{\frac{3}{4}} +1.5599i\epsilon\\ +0.0226i\epsilon^{\frac{5}{4}}+\mathcal{O}(\epsilon^{\frac{3}{2}}), &d=6-\epsilon\,.\\
    \end{cases}
\end{equation}
Pad\'e extrapolation gives $\braket{\bar{\sigma}}_{N=1}^{[1,2]}=-0.7646 i T^{\Delta_\sigma}$ in $d=2$, which differs from an exact result by $8.4\%$.

In $d=2$, effective central charge \eqref{eq:ceff} of $M(3,8)$ minimal model $c_\text{eff}(3,8)=\frac{3}{4}$. We can find thermal free energy density using \eqref{eq:freeenergydensity} and (\ref{freeEnergyExpansionRenorm}):
\begin{equation}
\begin{aligned}
\frac{f_{N=1}}{T^{6-\epsilon}}=
\begin{cases}
-\frac{\pi}{8}, & d=2\,,\\
-0.1312+0.0559\sqrt{\epsilon}-0.0934\,\epsilon+0.0868\,\epsilon^{\frac{5}{4}}\\
-0.0208\,\epsilon^{\frac{3}{2}}-0.0013\,\epsilon^{\frac{7}{4}}+\mathcal{O}(\epsilon^{2}),
& d=6-\epsilon\,.
\end{cases}
\end{aligned}
\end{equation}
Suprisingly, direct substitution of \(\epsilon=4\) into the truncated series up to \(\mathcal{O}(\epsilon)\) gives
\(f^{\mathcal{O}(\epsilon)}_{N=1}=-0.3931\,T^2\), within \(0.1\%\) of the exact \(M(3,8)\) value
\(f^{\text{exact}}_{M(3,8)}=-0.3926\,T^2\). This agreement supports the conjectured cubic two-field Lagrangian description of the \(M(3,8)_D\) minimal model \cite{Fei:2014xta,Klebanov:2022syt,Katsevich:2024jgq}.
We also perform two-sided Pad\'e extrapolations for the cubic \(N=1\) model imposing the \(d=2\) boundary condition (see Fig.~\ref{fig:N=1}). Table~\ref{tab:N=1} lists representative Pad\'e predictions for \(f_{N=1}/T^d\)  at \(d=3,4,5\).

\begin{table}[]
    \centering
    \begin{tabular}{|c|c|c|c|c|c|}
    \hline
        Dimension & $d=5$ & $d=4$ & $d=3$ & $d=2$\\\hline
        $f^{[2,3]}_{N=1}/T^d$ & $-0.1790$ & $-0.2416$ & $-0.3108$ & $-0.3927$ \\\hline
        $f^{[5,1]}_{N=1}/T^d$ & $-0.1686$ & $-0.2388$ & $-0.3144$ & $-0.3927$ \\\hline
        $f^{[4,2]}_{N=1}/T^d$ & $-0.1487$ & $-0.2109$ & $-0.2925$ & $-0.3927$\\\hline
        $f^{[6,1]}_{N=1}/T^d$ & $-0.1210$ & $-0.1649$ & $-0.2526$ & $-0.3927$ \\\hline
        $f^{\mathcal{O}(\epsilon)}_{N=1}/T^d$ & $-0.1688$ & $-0.2390$ & $-0.3147$ & $-0.3931$ \\\hline
    \end{tabular}
    \caption{Two-sided Pad\'e extrapolations and $\mathcal{O}(\epsilon)$ expansion for the thermal free energy $\frac{f_\text{YL}(d)}{T^d}$
    in the cubic $N=1$ theory for $d=3,4,5$. Note that $f^{[4,1]}_{N=1}(d)$ and $f^{[5,1]}_{N=1}(d)$ differ from each other less than by $0.02\%$ at range $2<d<6$.}
    \label{tab:N=1}
\end{table}

\begin{figure}
    \centering
    \includegraphics[width=0.6\linewidth]{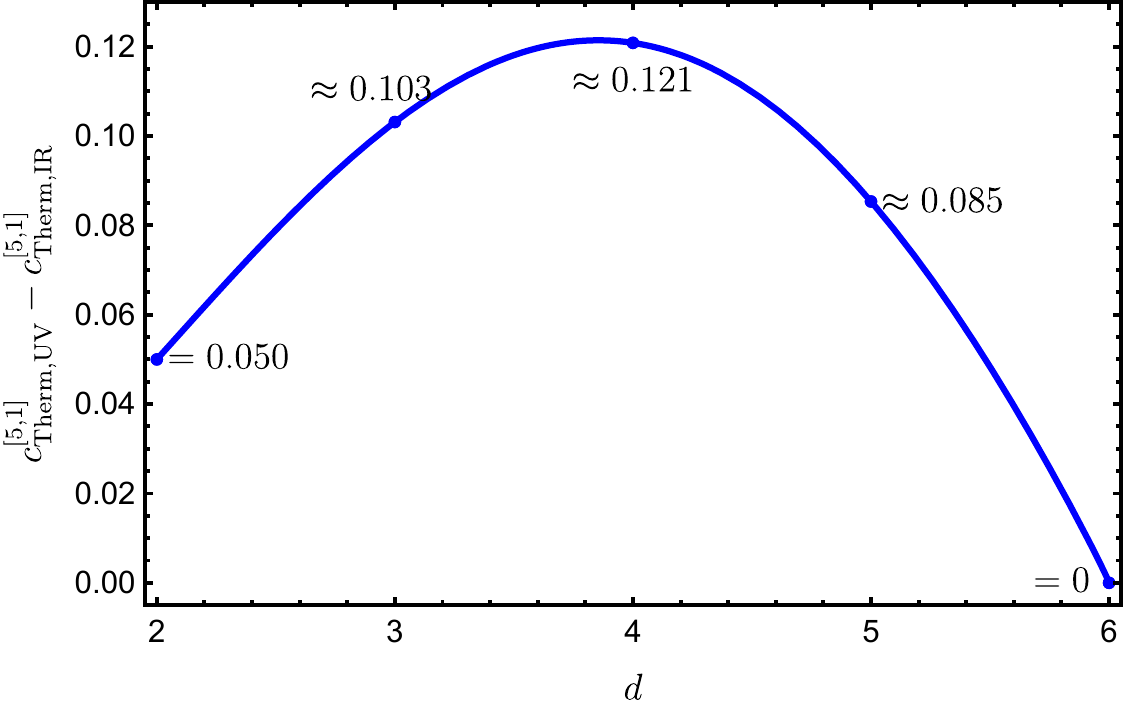}
    \caption{The difference between $c^{[5,1]}_\text{Therm,UV}(d)=2c^{[5,1]}_\text{Therm,YL}(d)$ and $c^{[5,1]}_\text{Therm,IR}(d)=c^{[5,1]}_{\text{Therm},N=1}(d)$ for the non-unitary flow from the two copies of YL theory to the non-trivial $N=1$ point.}
    \label{fig:ctheorem}
\end{figure}

Violation of the $F$-theorem corresponds to a violation of Zamolodchikov's $c$-theorem, since in $d=2$ sphere free energy is proportional to a central charge $F_{S^2}=\frac{\pi c}{6}$. In contrast, two-dimensional thermal free energy is proportional to effective central charge \eqref{eq:freeenergydensity} $f=-\frac{\pi c_\text{eff}}{6}T^2$, so we can expect that $c_\text{Therm}$-theorem is not violated for this non-unitary flow. Indeed, $c_\text{Therm,UV}(d)>c_\text{Therm,IR}(d)$ for all $d=2,3,4,5$ (see Figure \ref{fig:ctheorem}). Unfortunately, the $c_\text{Therm}$-theorem can not be a candidate for the role of the $F_\text{eff}$-theorem because it is violated in the well-known unitary three-dimensional flow from the quartic $O(N)$ model to the $N-1$ free Goldstone bosons \cite{Sachdev:1993pr,Chubukov:1993aau}. Thus, the Question about the $F_\text{eff}$-theorem remains open which would be interesting to resolve in the future.

\section*{Acknowledgments}

We thank Simone Giombi, Igor R. Klebanov, Alessio Miscioscia, Zimo Sun and Yifan Wang for valuable
discussions. F.K.P. gratefully acknowledges support from the Theoretical Sciences Visiting Program (TSVP) at the Okinawa Institute of Science and Technology (OIST), during which the final part of this research was conducted.

\appendix

\section{Free energy of free massive scalar field}\label{App:FreeEnergyMassive}
In this appendix, we review the small mass expansion of the thermal free energy density of a free scalar field \cite{Laine:2016hma}:
\begin{equation}
\begin{aligned}
    f_\text{free}(m^2)=\frac{1}{2\beta}\sum \limits_{n=-\infty}^{+\infty}\int_{\mathbb{R}^{d-1}}\frac{d^{d-1}k}{(2\pi)^{d-1}} \log{\left(\omega^2_n+k^2+m^2\right)}
\end{aligned}
\end{equation}
with $\omega_n=\tfrac{2\pi n}{\beta}$. For the massless case:
\begin{equation}\label{eq:maslessfreeen}
\begin{aligned}
    f_\text{free}(0)&=\frac{1}{2\beta}\sum \limits_{n=-\infty}^{\infty}\int_{\mathbb{R}^{d-1}}\frac{d^{d-1}k}{(2\pi)^{d-1}} \log{\left(\omega^2_n+k^2\right)}=-\frac{1}{2\beta}\sum_{n=-\infty}^\infty\int_0^\infty\frac{dt}{t}\int\frac{d^{d-1}k}{(2\pi)^{d-1}}e^{-t(\omega_n^2+k^2)}\\
    &=-\frac{\pi^{\frac{d-1}{2}}T^d}{2}\Gamma\left(\frac{1-d}{2}\right)\sum_{n=-\infty}^\infty n^{d-1}=-\frac{\Gamma(\frac{d}{2})\zeta(d)}{\pi^{\frac{d}{2}}}T^d\,.
\end{aligned}
\end{equation}
To simplify the computation of $f_\text{free}(m^2)$, we differentiate with respect to the mass squared to get
\begin{equation}\label{eq:FreeEnergyMassiveDerivative1}
    \Pi_0(m^2)\equiv2\frac{\partial f_\text{free}}{\partial m^2}=\frac{1}{\beta}\sum \limits_{n=-\infty}^{\infty}\int_{\mathbb{R}^{d-1}}\frac{d^{d-1}k}{(2\pi)^{d-1}}\frac{1}{\omega^2_n+k^2+m^2}\,.
\end{equation}
We note that for $\omega_n \neq 0$, the integrand can be expanded as a Taylor series in the parameter $m$. Subsequently, we can perform the summation over Matsubara modes and the integration over spatial momentum:
\begin{equation}\label{eq:FreeEnergyMassiveDerivative}
\begin{aligned}
    &\Pi_0(m^2)=\frac{1}{\beta}\int_{\mathbb{R}^{d-1}}\frac{d^{d-1}k}{(2\pi)^{d-1}}\frac{1}{k^2+m^2}+\frac{1}{\beta}\sum \limits_{n=1}^{\infty}\sum \limits_{L=0}^{\infty}\frac{\Gamma \left(\frac{3-d}{2}+L\right)}{2^{d-2}\pi^{\frac{d-1}{2}}\Gamma(L+1)}\frac{(-m^{2})^L}{\omega_n^{2L+3-d}}\\
    &=m^{d-3}T\frac{\Gamma\left(\frac{3-d}{2}\right)}{2^{d-1}\pi^{\frac{d-1}{2}}}+\frac{T^{d-2}}{2\pi^{\frac{5-d}{2}}}\sum \limits_{L=0}^{+\infty}\frac{(-1)^Lm^{2L}\Gamma\left(\frac{3-d}{2}+L\right)}{(2\pi T)^{2L}\Gamma(1+L)}\zeta(2L+3-d)\\
    &=\frac{T^{d-2}}{2\pi^{\frac{5-d}{2}}}\Gamma\left(\frac{3-d}{2}\right) \zeta(3-d)+\frac{m^{d-3}T}{2^{d-1}\pi^{\frac{d-1}{2}}}\Gamma\left(\frac{3-d}{2}\right)-\frac{m^2 T^{d-4}}{8\pi^{\frac{9-d}{2}}}\Gamma \left(\frac{5-d}{2}\right)\zeta (5-d)\\
    &+\frac{m^4 T^{d-6}}{64\pi^{\frac{13-d}{2}}}\Gamma\left(\frac{7-d}{2}\right)\zeta (7-d) +\mathcal{O}(m^6)\,.
\end{aligned}
\end{equation}
Integrating (\ref{eq:FreeEnergyMassiveDerivative}) with massless condition (\ref{eq:maslessfreeen}), we get that the free energy admits the following small $m$ expansion:
\begin{equation}\label{eq:FreeEnergyMassive}
\begin{aligned}
    f_\text{free}(m^2)&=-\frac{\Gamma\left(\frac{d}{2}\right)\zeta(d)}{\pi^{\frac{d}{2}}}T^d-\frac{m^{d-1} T }{2^{d}\pi ^{\frac{d-1}{2}}}\Gamma\left(\frac{1-d}{2}\right)\\
    &+\frac{m^2T^{d-2}}{4\pi^{\frac{5-d}{2}}}\sum_{L=0}^\infty\frac{(-1)^Lm^{2L}\Gamma(\frac{3-d}{2}+L)\zeta(3-d+2L)}{(2\pi T)^{2L}\Gamma(2+L)}\\
    &=-\frac{\Gamma\left(\frac{d}{2}\right)}{\pi^{\frac{d}{2}}}\zeta(d)T^d-\frac{m^{d-1} T }{2^{d}\pi ^{\frac{d-1}{2}}}\Gamma \left(\frac{1-d}{2}\right)+\frac{m^2 T^{d-2} }{4 \pi ^{\frac{5-d}{2}}}  \Gamma \left(\frac{3-d}{2}\right)\zeta (3-d)\\
    &-\frac{m^4 T^{d-4}}{32 \pi ^{\frac{9-d}{2}}} \Gamma \left(\frac{5-d}{2}\right)  \zeta (5-d)+\frac{ m^6 T^{d-6}}{384 \pi ^{\frac{13-d}{2}}}\Gamma \left(\frac{7-d}{2}\right) \zeta (7-d) +\mathcal{O}(m^8)\,.
\end{aligned}
\end{equation}
Note that the free energy has a $\frac{1}{\epsilon}$ pole near even dimensions $d=2n-\epsilon$, coming from vacuum energy of flat space-time.

\section{Perturbative solution of gap equation}
\label{App:solvevac}

In this section, we will perturbatively solve the gap equation \eqref{GapEquationO(N)} in the MS scheme. 
It is convenient to work with the dimensionless quantities, and thus we rescale $g_{i,0} \rightarrow \mu^{\frac{\epsilon}{2}}g_{i,0}$ and $v \rightarrow T^{2-\frac{\epsilon}{2}} v$. After that, the whole $T$ dependence drops out from the gap equation, and the analysis gets simplified. In the equation \eqref{GapEquationO(N)} the UV divergences come from two reasons. First reason is that the bare coupling constants $g_{i,0}$ contain the UV divergences and the second reason is that the equation itself contains UV divergences for any finite $g_{i,0}$ (see appendix~\ref{App:FreeEnergyMassive}), thus we get
\begin{equation}
    \begin{aligned}
    &-g_{2,0}v^2=\frac{\pi\left(N g_{1,0}+g_{2,0}\right)}{180}\left(1+\frac{A}{6}\epsilon\right)\\
    &+Ng_{1,0}\left(-\frac{g_{1,0}v}{48\pi}+\frac{\left(g_{1,0} v\right)^{\frac32}}{24 \pi ^2}+\frac{\left(g_{1,0}v\right)^2}{64\pi^3} \left(\frac{1}{\epsilon}-\frac{1}{2}\log\left(\frac{4\pi e^{-\gamma_E}T^2}{\mu^2}\right)\right)\right)\\
    &+g_{2,0}\left(-\frac{g_{2,0} v}{48\pi}+\frac{\left(g_{2,0} v\right)^\frac32}{24 \pi ^2}+\frac{\left(g_{2,0} v\right)^2}{64\pi^3}\left(\frac{1}{\epsilon}-\frac{1}{2}\log\left(\frac{4\pi e^{-\gamma_E}T^2}{\mu^2}\right)\right)\right)+\mathcal{O}\left(g^4_{i,0}v^3\right)\,,
\end{aligned}
\end{equation}
where
\begin{gather}\label{Adef}
    A=8-3\log\left(4\pi  e^{\gamma_E}\right)+720\zeta'(-3)\,.
\end{gather}
To get a perturbative solution of the above equation, we have to deal with the ratios $\tfrac{g_{1,0}}{g_{2,0}}$. For that, we notice that at a fixed point, we have that $g_i \sim \sqrt{\epsilon}$. That is why we rescale $g_{i} \rightarrow t  g_{i}$, and then expand our gap equation in parameter $t$:
\begin{equation}
\begin{aligned}
    &-v^2=\frac{\pi\left(Ng_{1}+g_{2}\right) }{180 g_{2}}\left(1+ \epsilon\frac{A}{6}\right)-vt \frac{N g^2_{1}+g^2_2}{48\pi g_2}\\
    &+t^{\frac32}\frac{N g_1\left(g_{1} v\right)^{\frac32}+g_2\left(g_{2} v\right)^{\frac32}}{24 \pi ^2 g_2}+t^2\left(v^2\frac{Ng^3_1+g^3_2}{64\pi^3g_2}\left(\frac{1}{\epsilon}-\frac{1}{2}\log\left(\frac{4\pi e^{-\gamma_E}T^2}{\mu^2}\right)\right)\right.\\
    &\left.+\frac{ N g_1  \left(6Ng_1^3-(N+4)g_1^2 g_2 -6g_1 g_2^2+5 g_2^3\right)}{69120 \pi ^2  g_2^{2} }\left(\frac{1}{\epsilon}+\frac{A}{6}\right)\right)+\mathcal{O}(t^3)\,.
\end{aligned}
\end{equation}
With this equation, we solve for $v$ in the following form in $t$ (recovering $T$ dependents): 
\begin{equation}\label{vSolExpansionIn_t}
    \frac{v}{T^{2-\frac{\epsilon}{2}}}=v_0+  v_1 t +  v_{\frac{3}{2}} t^{\frac{3}{2}} + v_2  t^2 +v_{\frac{5}{2}} t^{\frac{5}{2}} +\mathcal{O}( \epsilon t, t^3,\epsilon^2)\,,
\end{equation}
where
\begin{equation}\label{eq:vacsol}
    \begin{aligned}
    &v_0=-\frac{i}{6}\sqrt{\frac{\pi(Ng_1+g_2)}{5g_2}}\left(1+\frac{A}{12}\epsilon\right)\,,\\
    &v_1=\frac{Ng^2_{1}+g^2_{2}}{96\pi g_{2}},\quad v_{\frac{3}{2}}=\left(\frac{Ng_1+g_2}{5g_2}\right)^{\frac{1}{4}}\frac{N\left(-i g_1\right)^{\frac{5}{2}}+\left(-i g_2\right)^{\frac{5}{2}}}{48\sqrt{6}\pi^{\frac{7}{4}}g_{2}}\,,\\
    &v_2=i\sqrt{\frac{5g_2}{Ng_1+g_2}}\left(\frac{10Ng_1^3 +(N g_1 +6 g_2) \left(Ng_1^2+g_2^2\right)}{23040\pi^{\frac{5}{2}}g_2\epsilon}+\frac{\left(Ng_1^2+g_2^2\right)^2}{3072\pi^{\frac{5}{2}}g_2^2}\right.\\
    &\left.+\frac{\left(10 Ng_1^3 +(Ng_1 +6 g_2) \left(Ng_1^2+g_2^2\right)\right)}{276480\pi^{\frac{3}{2}}g_2}A-\frac{(Ng_1+g_2)(N g_1^3+g_2^3)\log{\left(\frac{4\pi e^{-\gamma_E}T^2}{\mu^2}\right)}}{7680\pi^{\frac{5}{2}}g_2^2}\right)\,,\\
    &v_{\frac{5}{2}}= i\left(\frac{5g_2}{Ng_1+g_2}\right)^{\frac{1}{4}}\frac{(Ng_1^2+g_2^2)(N(-i g_1)^{\frac{5}{2}}+\left(-i g_2\right)^{\frac{5}{2}})}{512\sqrt{6}\pi^{\frac{13}{4}}g^2_2}\,.
\end{aligned}
\end{equation}
We set $t=1$ after performing the expansion in \eqref{vSolExpansionIn_t} with the above solution. Indeed, the parameter $t$ is a formal expansion parameter, and each term has already been collected according to its total power in the couplings $g_i$.

Let us note that there are two imaginary solutions for $g_{i}$ at a fixed point related by complex conjugation, as well as two solutions for $v=\pm i r_0+\mathcal{O}(g_i)$, where $r_0$ is real. We must choose these signs $g_i$ and $v$ such that the resulting thermal mass for both scalar fields after renormalization is positive $m_i^2 >0$, which enforces the first sign in the above equation and determines the consecutive expansion.

\section{Small $m$ expansion of $\mathcal{I}_2$}\label{App:I_2}
In this section, we study a small-mass expansion of the following sum-integral
\begin{equation}\label{I^{(1)}_2}
   \mathcal{I}_2(m^2_1,m^2_2,m^2_3)= \frac{1}{\beta^2}\sum \limits_{n_p,n_q}\int\frac{d^{d-1}pd^{d-1}q}{(2\pi)^{2(d-1)}}\frac{1}{(P^2+m^2_1)(Q^2+m^2_2)((P+Q)^2+m^2_3)}
\end{equation}
up to cubic order in masses $m_a$ with $a=1,2,3$, where for brevity we also introduce notation $P = (p_0, p)$, $Q = (q_0, q)$, and where we assume that $p_0=\tfrac{2\pi n_p}{\beta}$ and $q_0=\tfrac{2\pi n_q}{\beta}$. The same integral has already been considered in Appendix F of \cite{Arnold:1994ps} in $d=4-\epsilon$. The main complication is that a naive expansion in small masses $m_i^2$ is contaminated by IR divergences, thus making it invalid. It is connected due to the presence of the zero Matsubara mode. To isolate and regulate these divergences, we split the integral $\mathcal{I}_2(m^2_1,m^2_2,m^2_3)$ into three parts:
\begin{equation}
    \mathcal{I}_2(m^2_1,m^2_2,m^2_3)=\mathcal{I}^{(1)}_2+\mathcal{I}^{(2)}_2 +\mathcal{I}^{(3)}_2
\end{equation}
with
\begin{equation}
\begin{aligned}
    \mathcal{I}^{(1)}_2&= \frac{1}{\beta^2}\sum \limits_{n_p,n_q}\int\frac{d^{d-1}pd^{d-1}q}{(2\pi)^{2(d-1)}}\frac{1}{(P^2+m^2_1)(Q^2+m^2_2)((P+Q)^2+m^2_3)}\\
    &-\frac{1}{\beta^2}\sum \limits_{n_p,n_q}\int\frac{d^{d-1}pd^{d-1}q}{(2\pi)^{2(d-1)}}\frac{1}{P^2Q^2(P+Q)^2}+\frac{m_1^2}{\beta^2}\sum \limits_{n_p,n_q}\int\frac{d^{d-1}pd^{d-1}q}{(2\pi)^{2(d-1)}}\frac{1}{P^4Q^2(P+Q)^2}\\
    &+\frac{m_2^2}{\beta^2}\sum \limits_{n_p,n_q}\int\frac{d^{d-1}pd^{d-1}q}{(2\pi)^{2(d-1)}}\frac{1}{P^2Q^4(P+Q)^2}+\frac{m_3^2}{\beta^2}\sum \limits_{n_p,n_q}\int\frac{d^{d-1}pd^{d-1}q}{(2\pi)^{2(d-1)}}\frac{1}{P^2Q^2(P+Q)^4}\,,
\end{aligned}
\end{equation}
as well as
\begin{equation}
\begin{aligned}
    \mathcal{I}^{(2)}_2&=\frac{1}{\beta^2}\sum \limits_{n_p,n_q}\int\frac{d^{d-1}pd^{d-1}q}{(2\pi)^{2(d-1)}}\frac{1}{P^2Q^2(P+Q)^2}\,,\\
    \mathcal{I}^{(3)}_2&=- \sum\limits_{a=1}^{3} \frac{m_a^2}{\beta^2}\sum \limits_{n_p,n_q}\int\frac{d^{d-1}pd^{d-1}q}{(2\pi)^{2(d-1)}}\frac{1}{P^4Q^2(P+Q)^2}\,.
\end{aligned}
\end{equation}
Let us immediately point out that the term $\mathcal{I}^{(2)}_2$ vanishes. Indeed, summing over $n_p$ and $n_q$ we get the following expression
\begin{equation}
\begin{aligned}
   \mathcal{I}^{(2)}_2&= \frac{1}{\beta^2}\sum \limits_{n_p,n_q}\int\frac{d^{d-1}pd^{d-1}q}{(2\pi)^{2(d-1)}}\frac{1}{P^2Q^2(P+Q)^2}\\
   &=-\frac{3 }{16}\int\frac{d^{d-1}pd^{d-1}q}{(2\pi)^{2(d-1)}}\frac{1}{|p||q|}\frac{\coth{\left(\frac{\beta|p|}{2} \right)}\coth{\left(\frac{\beta|q|}{2} \right)}-1}{(|p|-|q|)^2-(p+q)^2}\frac{\left(|p|^2+|q|^2-(p+q)^2\right)}{(|p|+|q|)^2-(p+q)^2}\\
   &=-\frac{3}{32}\int\frac{d^{d-1}pd^{d-1}q}{(2\pi)^{2(d-1)}}\frac{\left(p\cdot q\right)}{|p||q|}\frac{\coth{\left(\frac{\beta|p|}{2} \right)}\coth{\left(\frac{\beta|q|}{2} \right)}-1}{(|p||q|)^2-(p\cdot q)^2}=0\,,
\end{aligned}
\end{equation}
since its odd in $p$ and  $q$. Note that after summation over $n_1$ and $n_2$, we will also get a temperature-independent term, which is removed by the flat space-time renormalization.

Now, we start with the analysis of $\mathcal{I}^{(1)}_2$. For that, we note that we can expand the sum in the following way
\begin{equation}\label{doublesum}
\begin{aligned}
    \sum\limits_{n_p,n_q}I(n^2_p,n^2_q,(n_p+n_q)^2)&=I(0,0,0)+\sum_{\substack{ n\neq 0}}\left(I(n^2,0,n^2)+I(0,n^2,n^2)+I(n^2,n^2,0)\right)\\
    &+\sum_{\substack{ n_p\neq 0,\; n_q\neq 0,\; \\n_p+n_q\neq 0}}I(n^2_p,n^2_q,(n_p+n_q)^2)\,.
\end{aligned}
\end{equation}
Since the last term contains no zero Matsubara modes, its expansion in the masses is well defined. 
Applying \eqref{doublesum} term by term to $\mathcal{I}^{(1)}_2$, we find that the double sums over $n_p$ and $n_q$ cancel through quadratic order in the masses. 
Thus, only single sums remain, which can be evaluated explicitly. 
In these sums, we again expand only those masses that appear together with nonzero Matsubara frequencies. 
Keeping terms up to $\mathcal{O}(m_a^2)$, we obtain:
\begin{equation}
\begin{aligned}
    \mathcal{I}^{(1)}_2&=\frac{1}{\beta^2}\int\frac{d^{d-1}pd^{d-1}q}{(2\pi)^{2(d-1)}}\frac{1}{(p^2+m^2_1)(q^2+m^2_2)((p+q)^2+m^2_3)}\\
    &-\frac{1}{\beta^2}\int\frac{d^{d-1}pd^{d-1}q}{(2\pi)^{2(d-1)}}\frac{1}{p^2q^2(p+q)^2}+\sum\limits_{a=1}^{3}\frac{m^2_a}{\beta^2}\int\frac{d^{d-1}pd^{d-1}q}{(2\pi)^{2(d-1)}}\frac{1}{p^4q^2(p+q)^2}\\
    &-\sum\limits_{a=1}^{3}\frac{m^2_a}{\beta^2}\sum_{\substack{ n_q\neq 0}}\int\frac{d^{d-1}pd^{d-1}q}{(2\pi)^{2(d-1)}}\frac{1}{p^2(p^2+m^2_a)}\frac{1}{(q^2+q^2_0)((p+q)^2+q^2_0)}\\
    &+\sum\limits_{a=1}^{3}\frac{m^2_a}{\beta^2}\sum_{\substack{ n_q\neq 0}}\int\frac{d^{d-1}pd^{d-1}q}{(2\pi)^{2(d-1)}}\frac{1}{p^4(q^2+q^2_0)((p+q)^2+q^2_0)}+\mathcal{O}(m_1^{n_1}m_2^{n_2}m_3^{n_3})
\end{aligned}
\end{equation}
with $n_1+n_2+n_3=4$. First of all, we see that the integral in the first line is of order $m^{2d-8}_1$, and thus can be neglected.  The integrals in the second line will vanish in dimensional regularization, and thus we will get the following form:
\begin{equation}\label{I2(1)withsum}
\begin{aligned}
    \mathcal{I}^{(1)}_2&= -\sum_{a=1}^{3}\frac{m^2_a}{\beta^2}\sum_{\substack{ n_q\neq 0}}\int\frac{d^{d-1}pd^{d-1}q}{(2\pi)^{2(d-1)}}\frac{1}{p^2(p^2+m^2_a)}\frac{1}{(q^2+q^2_0)((p+q)^2+q^2_0)}\\
    &+\sum_{a=1}^{3}\frac{m^2_a}{\beta^2} \sum_{\substack{n_q\neq 0}}\int\frac{d^{d-1}pd^{d-1}q}{(2\pi)^{2(d-1)}}\frac{1}{p^4(q^2+q^2_0)((p+q)^2+q^2_0)}+\mathcal{O}(m_1^{n_1}m_2^{n_2}m_3^{n_3})\,,
\end{aligned}
\end{equation}
where $n_1+n_2+n_3=4$. The summation over Matsubara modes $n_q$ can be taken explicitly, by adding and subtracting the $n_q=0$ mode, and performing the summation over Matsubara frequencies as well as taking trivial integrals over $q$ using that:
\begin{equation}
\begin{aligned}
    &\frac{1}{\beta}\sum_{\substack{ n_q\neq 0}}
    \int\frac{d^{d-1}q}{(2\pi)^{d-1}}\frac{1}{(q^2+q^2_0)((p+q)^2+q^2_0)}= -\frac{1}{\beta p^{5-d}}\frac{2^{5-2d}\pi^{2-\frac{d}{2}}}{\Gamma(\frac{d}{2}-1)\cos\left(\frac{\pi  d}{2}\right)}\\
    &+\frac{1}{p^{4-d}}\frac{2^{-d} \pi ^{-\frac{d}{2}} \Gamma \left(2-\frac{d}{2}\right) \Gamma \left(\frac{d}{2}-1\right)^2}{\Gamma (d-2)}+\int \frac{d^{d-1}q}{(2\pi)^{d-1}} \frac{1}{( p + q)^2 - q^2}\left(\frac{n(\beta|q|)}{|q|} - \frac{n(\beta|q + p|)}{|q + p|}\right)\,,
\end{aligned}
\end{equation}
where we have defined $n(x) = \frac{1}{e^{x} - 1}$. That leads to the following result 
\begin{equation}
\begin{aligned}
    \mathcal{I}^{(1)}_2 &= -\sum\limits_{a=1}^{3}\frac{m^2_a}{\beta}
    \int\frac{d^{d-1}pd^{d-1}q}{(2\pi)^{2(d-1)}}\frac{1}{p^2(p^2+m^2_a)}\frac{1}{(p + q)^2 - q^2}\left(\frac{n(\beta|q|)}{|q|}-\frac{n(\beta|q+p|)}{|q+p|}\right)\\
    &+\sum\limits_{a=1}^{3}\frac{m^2_a}{\beta}\int\frac{d^{d-1}pd^{d-1}q}{(2\pi)^{2(d-1)}}\frac{1}{p^4}\frac{1}{( p + q)^2 - q^2}\left(\frac{n(\beta|q|)}{|q|}-\frac{n(\beta|q+p|)}{|q+p|}\right)+\mathcal{O}(m^{2d-8}_a)\,,
\end{aligned}
\end{equation}
where again we neglected terms that are higher-order in the masses $m_a$. Finally, let us consider the last term $ \mathcal{I}^{(3)}_2$. We can perform the following expansion
\begin{equation}\label{eqb.12}
\begin{aligned}
    \mathcal{I}^{(3)}_2&= -\sum\limits_{a=1}^{3}\frac{m_a^2}{\beta}\left(\int\frac{d^{d-1}p}{(2\pi)^{d-1}}\frac{1}{p^4}\int \frac{d^{d-1}q}{(2\pi)^{d-1}}\frac{1}{p^2+2q\cdot p}\left(\frac{n(\beta|q|)}{|q|}-\frac{n(\beta|q+p|)}{|q+p|}\right)\right.\\
    &+\sum_{\substack{ n_p\neq 0}}\int\frac{d^{d-1}p}{(2\pi)^{d-1}}\frac{1}{P^{8-d}}\frac{ \Gamma \left(2-\frac{d}{2}\right) \Gamma \left(\frac{d}{2}-1\right)^2}{(4\pi)^{d/2}\Gamma (d-2)}\\
    &\left.+2\sum_{\substack{  n_p\neq 0}}\int\frac{d^{d-1}p}{(2\pi)^{d-1}}\frac{1}{P^4}\int \frac{d^{d-1}q}{(2\pi)^{d-1}}\frac{n(\beta|q|)}{|q|}\frac{P^2+2p \cdot q}{(P^2+2q\cdot p)^2+4 q^2 p^2_0}\right)\,,
\end{aligned}
\end{equation}
where in the last two lines we used that for $p_0 \neq 0$
\begin{equation}\label{PiThermal}
\begin{aligned}
    &\frac{1}{\beta}\sum \limits_{n_q}\int\frac{d^{d-1}q}{(2\pi)^{d-1}}\frac{1}{ Q^2(P+Q)^2}=P^{d-4}\frac{\Gamma\left(2-\frac{d}{2}\right)\Gamma \left(\frac{d}{2}-1\right)^2}{(4\pi)^{\frac{d}{2}}\Gamma (d-2)}\\
    &+2\int \frac{d^{d-1}q}{(2\pi)^{d-1}}\frac{n(\beta|q|)}{|q|}\frac{P^2+2p \cdot q}{(P^2+2q\cdot p)^2+4 q^2 p^2_0}\,.
\end{aligned}
\end{equation}
Note that the IR divergent term in the first line of \eqref{eqb.12} will be exactly canceled from contribution from $\mathcal{I}^{(1)}_2$ as expected.
In order to extract UV divergences of integrals in the second and third lines in \eqref{eqb.12}, we need to subtract leading large $P$ behavior which can be easily deduced from \eqref{PiThermal}:
\begin{equation}
\begin{aligned}
     &\lim_{P\to\infty}\frac{1}{\beta}\sum \limits_{n_q}\int\frac{d^{d-1}q}{(2\pi)^{d-1}}\frac{1}{ Q^2(P+Q)^2}=P^{d-4}\frac{\Gamma \left(2-\frac{d}{2}\right) \Gamma \left(\frac{d}{2}-1\right)^2}{(4\pi)^{\frac{d}{2}}\Gamma (d-2)}\\
     &+\frac{T^{d-2}}{P^2}\frac{\Gamma\left(\frac{d-2}{2}\right)\zeta(d-2)}{\pi^{\frac{d}{2}}}+\mathcal{O}\left(\frac{1}{P^4}\right)
\end{aligned}
\end{equation}
with the result
\begin{equation}
    \begin{aligned}
    \mathcal{I}^{(3)}_2
    &=-\sum\limits_{a=1}^{3}\frac{m_a^2}{\beta}\Bigg(\int\frac{d^{d-1}p}{(2\pi)^{d-1}}\frac{1}{p^4}\int\frac{d^{d-1}q}{(2\pi)^{d-1}}\frac{1}{p^2+2q\cdot p}\left(\frac{n(\beta|q|)}{|q|}-\frac{n(\beta|q+p|)}{|q+p|}\right)\\
    &+\sum_{\substack{n_p\neq 0}}\int\frac{d^{d-1}p}{(2\pi)^{d-1}}\frac{1}{P^{8-d}}\frac{ \Gamma \left(2-\frac{d}{2}\right) \Gamma \left(\frac{d}{2}-1\right)^2}{(4\pi)^{d/2}\Gamma(d-2)}\\
    &+\sum_{\substack{ n_p\neq 0}}\int\frac{d^{d-1}p}{(2\pi)^{d-1}}\frac{T^{d-2}}{P^6}\frac{1}{\pi^{d/2}}\Gamma\left(\frac{d-2}{2}\right)\zeta(d-2)\\
    &-4\sum_{\substack{ n_p\neq 0}}\int\frac{d^{d-1}p}{(2\pi)^{d-1}}\frac{1}{P^6}\int \frac{d^{d-1}q}{(2\pi)^{d-1}}\frac{n(\beta|q|)}{|q|}\frac{p\cdot q\left(P^2+2p \cdot q\right)+2q^2 p^2_0}{(P^2+2q\cdot p)^2+4 q^2 p^2_0}\Bigg)\,.
    \end{aligned}
\end{equation}
Putting everything together and expanding integrals in $\epsilon$, we get that
\begin{equation}
      \mathcal{I}_2(m^2_1,m^2_2,m^2_3) = -T^{2d-8}\sum\limits_{a=1}^{3} m_a^2\mathcal{S}^{a}_2+\mathcal{O}(m_1^{n_1}m_2^{n_2}m_3^{n_3})\,,
\end{equation}
with $n_1+n_2+n_3=4$, and where we defined $\mathcal{S}^{a}_2$ as
\begin{equation}\label{S2a}
    \begin{aligned}
        \mathcal{S}^{a}_2 &=\frac{1}{6912 \pi^2 \epsilon} +\frac{ \pi ^4 (15\gamma_E -10-18 \log (2)-3 \log (\pi ))-1080 \zeta '(4)}{103680 \pi ^6}\\
        &+\int\frac{d^{5}p}{(2\pi)^{5}}\frac{1}{p^2 (p^2 + (m_a \beta)^2)}\int \frac{d^{5}q}{(2\pi)^{5}}\frac{1}{(p + q)^2 - q^2}\left(\frac{n(|q|)}{|q|}-\frac{n(|q+p|)}{|q+p|}\right)\\
        &-4\sum_{\substack{ p_0\neq 0}}\int\frac{d^{5}p}{(2\pi)^{5}}\frac{1}{P^6}\int \frac{d^{5}q}{(2\pi)^{5}}\frac{n(|q|)}{|q|}\frac{p\cdot q\left(P^2+2p \cdot q\right)+2q^2 p^2_0}{(P^2+2q\cdot p)^2+4 q^2 p^2_0}\,,
    \end{aligned}
\end{equation}
where in the first line we only keep terms to order $\epsilon^0$. The integral in the second line is evaluated in Appendix \ref{App: J_1}, and the integral in the last line of \eqref{S2a} evaluates numerically to:
\begin{equation}
   \mathcal{J}_2\equiv\sum_{\substack{ n_p\neq 0}}\int\frac{d^{5}p}{(2\pi)^{5}}\frac{1}{P^6}\int \frac{d^{5}q}{(2\pi)^{5}}\frac{n(|q|)}{|q|}\frac{p\cdot q\left(P^2+2p \cdot q\right)+2q^2 p^2_0}{(P^2+2q\cdot p)^2+4 q^2 p^2_0} =3.3033\cdot10^{-7}\,,
\end{equation}
which will give the final answer
\begin{equation}\label{I2FinalEpsExpansion}
    \mathcal{I}_2(m^2_1,m^2_2,m^2_3)= -T^{2d-8}\sum\limits_{a=1}^{3}m^2_a\left(\frac{1}{6912 \pi^2 \epsilon}+b-\frac{m_a\beta}{1152\pi^3}\right)+\mathcal{O}(m_1^{n_1}m_2^{n_2}m_3^{n_3})
\end{equation}
with $n_1+n_2+n_3=4$, and constant $b$ is given by
\begin{equation}\label{b_numericalvalue}
    b=\frac{ \pi ^4 (15\gamma_E -10-18 \log (2)-3 \log (\pi ))-1080 \zeta '(4)}{103680 \pi ^6}+\frac{\zeta(3)}{192\pi^4}-4\mathcal{J}_2=4.68376\cdot10^{-5}\,.
\end{equation}
\subsection{Integral $\mathcal{J}_1$}\label{App: J_1}
In this section, we provide a small mass $M=\beta m_a$ expansion of the following integral
\begin{equation}
\begin{aligned}
    \mathcal{J}_1&\equiv\int\frac{d^{5}p}{(2\pi)^{5}}\frac{1}{p^2(p^2+M^2)}\int \frac{d^{5}q}{(2\pi)^{5}}\frac{1}{(p+q)^2 - q^2}\left(\frac{n(|q|)}{|q|}-\frac{n(|p+q|)}{|p+q|}\right)\,,
\end{aligned}
\end{equation}
and show that this expression is not analytical in mass $M$. Indeed, by introducing $g(x)=\frac{n(\sqrt{x})}{\sqrt{x}}$, we observe that it can be represented in the following form: 
\begin{equation}
\begin{aligned}
    \mathcal{J}_1=-\int_0^1dt\int\frac{d^5p}{(2\pi)^5}\frac{1}{p^2(p^2+M^2)}\int\frac{d^5k}{(2\pi)^5}g'(k^2+t(1-t)p^2)\,,
\end{aligned}
\end{equation}
where we perform a change of variables $k=q+tp$. Next, by introducing $u=k^2+t(1-t)p^2$ the integral over $p$ can be rewritten as:
\begin{equation}
    \int\frac{d^5p}{(2\pi)^5}\frac{g'(k^2+t(1-t)p^2)}{p^2(p^2+M^2)}
    =\frac{1}{24\pi^3}\int_{k^2}^\infty\frac{du}{\sqrt{t(1-t)}}\frac{\sqrt{u-k^2}g'(u)}{u-k^2+t(1-t)M^2}\,.
\end{equation}
After exchanging the order of integration over $k$ and $u$, the $k$ integral is performed first. The resulting $u$ integral is simplified via integration by parts, and the $t$ integral can then be carried out explicitly. This procedure yields the following representation for $\mathcal{J}_1$:
\begin{equation}
\begin{aligned}
    &\mathcal{J}_1=\frac{1}{192\pi^5}\int_0^\infty dr\left(r^2+\frac{M^2}{4}\right)\arctan\left(\frac{2r}{M}\right)n(r)-\frac{M}{2304\pi^3}\\
    &=\frac{\zeta (3)}{192 \pi ^4}-\frac{M}{1152 \pi ^3}+\frac{1}{768 \pi ^5}\int \limits_0^{+\infty}dr\left(\left(M^2+4 r^2\right)\arctan\left(\frac{2 r}{M}\right)-2\pi  r^2+2 r M\right)n(r)\,.
\end{aligned}
\end{equation}
As we show below, the remaining integral is of order $M^2$. This representation is not suited for naive small mass expansion as it is non-analytic, as we show below. To find such an expansion, we need to introduce an intermediate scale $R$, such that $M\ll R\ll1$. Then we can divide the integral into two parts:
\begin{equation}
\int_0^\infty=\int_0^R+\int_R^\infty\,.
\end{equation}
The first integral $\int_0^R$ after rescaling has the following form:
\begin{equation}
\begin{aligned}
   \frac{M^3}{768 \pi ^5}\int_0^{\frac{R}{M}}dy\left(\left(1+4 y^2\right)\arctan\left(2 y\right)-2\pi  y^2+2 y \right)n(M y)\,.
\end{aligned}
\end{equation}
We can expand $n(My)$ in small parameter $M y$, and then integrate term by term with result:
\begin{equation}
\begin{aligned}
    &\frac{M^3}{768 \pi ^5}\int_0^{\frac{R}{M}}dy((1+4y^2) \arctan\left(2 y\right)-2\pi  y^2+2 y )n(M y)\\
    &=\frac{M^2 \left(1+2 \log \left(\frac{2R}{M}\right)\right)}{3072 \pi ^4}+\mathcal{O}\left(M^3,RM^2,\frac{M^3}{R}\right)\,.
\end{aligned}
\end{equation}
In the second integral, $\int_R^\infty$, we use that $\tfrac{R}{M}\gg 1$, so the integration region lies in the large-$y$ regime. We therefore expand the prefactor of $n(My)$ for large $y$ and subsequently perform the integration term by term:
\begin{equation}
\begin{aligned}
    &\frac{M^3}{768 \pi ^5}\int_{\frac{R}{M}}^{+\infty}dy\left(\left(1+4 y^2\right)\arctan\left(2 y\right)-2\pi  y^2+2 y \right)n(M y)\\
    &=-\frac{M^2 \log (R)}{1536 \pi ^4}+\mathcal{O}(M^3,R M^2)\,.
\end{aligned}
\end{equation}
Putting everything together, we get that
\begin{equation}
    \mathcal{J}_1=\frac{\zeta (3)}{192 \pi ^4}-\frac{M}{1152 \pi ^3}+\frac{ M^2(1+\log{4})}{3072 \pi ^4}-\frac{M^2\log M}{1536\pi^4}+\mathcal{O}(M^3)\,.
\end{equation}
where terms of the form $R^n M^2$ are expected to drop off since the integral is $R$ independent. We also have checked numerically, that this expansion is valid for the small $M$ expansion.

\section{Evaluation of $K_T(m^2_1,m^2_2;p_0,p)$}
\label{APP:KTIntegral}
We consider the integral
\begin{equation}
    K_T(m_1^2,m_2^2;p_0,p)=\frac{1}{\beta}\sum_{n_q}\int\frac{d^{d-1}q}{(2\pi)^{d-1}}\frac{1}{(q_0^2+q^2+m_1^2)((q_0+p_0)^2+(q+p)^2+m_2^2)}\,.
\end{equation}

A complete closed-form evaluation of this diagram is technically involved and unnecessary for the present analysis. Instead, we focus on two specific features: its ultraviolet (UV) divergent part and its small-mass expansion at $p_0=0$, which are sufficient for determining thermal masses perturbatively.

The UV divergence is independent of temperature and can therefore be extracted by taking the zero-temperature limit. In this limit, the Matsubara sum becomes a continuous Euclidean energy integral, and the diagram reduces to a standard vacuum loop integral. Evaluating it using dimensional regularization yields
\begin{equation}
\label{KTDivPart}
\left[K_T(m_1^2,m_2^2;p_0,p)\right]_{\rm div}
= -\frac{1}{(4\pi)^3 \epsilon}
\left(\frac{p_0^2+p^2}{3} + m_1^2 + m_2^2\right)\,.
\end{equation}

To determine the thermal masses perturbatively, we require the behavior of $K_T$ at $p_0=0$ expanded for small $p^2$ and small masses. We therefore consider
\begin{equation}
\begin{aligned}
K_T(m_1^2,m_2^2;0,p)&=\frac{1}{\beta}\sum_{n_q}\int\frac{d^{d-1}q}{(2\pi)^{d-1}}\frac{1}{\big(q_0^2+q^2+m_1^2\big)\big(q_0^2+(q+p)^2+m_2^2\big)}\\
&= K^{(n_q=0)}_T(m_1^2,m_2^2;0,p)+ K^{(n_q \neq 0)}_T(m_1^2,m_2^2;0,p)\,,
\end{aligned}
\end{equation}
where the zero-mode and non-zero Matsubara contributions have been separated for subsequent expansion. The contribution of the non-zero Matsubara modes is analytic in both $m_i^2$ and $p^2$. Using Feynman parametrization, one obtains
\begin{equation}
\begin{aligned}
K^{(n_q \neq 0)}_T(m_1^2,m_2^2;0,p)
=\frac{2\Gamma\left(\frac{5-d}{2}\right)}{(4\pi)^{\frac{d-1}{2}}}T \int_{0}^{1}dx \sum_{n_q=1}^{\infty}(q_0^2 + x(1-x)p^2 + x m_1^2 + (1-x)m_2^2)^{\frac{d-5}{2}}\\
=\frac{T^{d-4}\Gamma\left(\frac{5-d}{2}\right)\zeta(5-d)}{8\pi^{\frac{9-d}{2}}}
-\frac{T^{d-6}}{(4\pi)^3}
\left(\frac{p^2}{3}+m_1^2+m_2^2\right)
\frac{\zeta(7-d)\Gamma\!\left(\frac{7-d}{2}\right)}{\pi^{\frac{7-d}{2}}}
+\mathcal{O}\!\bigl(m_1^{n_1} m_2^{n_2} p^{n_3}\bigr)\,,
\end{aligned}
\end{equation}
where $n_1+n_2+n_3=4$. 

The zero-mode contribution is more subtle, as it is non-analytic in the small-mass expansion. It is given by
\begin{equation}
\begin{aligned}
K^{(n_q=0)}_T(m_1^2,m_2^2;0,p)= \frac{1}{\beta}\int\frac{d^{d-1}q}{(2\pi)^{d-1}}\frac{1}{(q^2+m_1^2)((q+p)^2+m_2^2)}\\
= \frac{T}{(4\pi)^{\frac{d-1}{2}}}\Gamma\!\left(\frac{5-d}{2}\right)
\int_{0}^{1}\!dx(x(1-x)p^2 + x m_1^2 + (1-x)m_2^2)^{\frac{d-5}{2}}\,.
\end{aligned}
\end{equation}
Specializing to $d=6$, we can evaluate the integral exactly and obtain
\begin{equation}
\begin{aligned}
    &K^{(n_q=0)}_T(m_1^2,m_2^2;0,p)= -\frac{T(m_1+m_2)\big(p^2-(m_1-m_2)^2\big)}{64\pi^2 p^2}\\
    &-\frac{T\big(p^2+(m_1+m_2)^2\big)\big(p^2+(m_1-m_2)^2\big)}{64\pi^2 p^3}
    \arctan\!\left(\frac{p}{m_1+m_2}\right)= T\, m_1\, h\!\left(\frac{p}{m_1},\frac{m_2}{m_1}\right),
\end{aligned}
\end{equation}
where the last equality makes explicit that the overall scaling is set by $m_1$, and $h(x,y)$ is a dimensionless function of the ratios $x=p/m_1$ and $y=m_2/m_1$.

\bibliographystyle{JHEP}
\bibliography{refs}

\end{document}